-
\documentclass[a4paper,fleqn,usenatbib]{mnras}

\usepackage{newtxtext,newtxmath}

\usepackage[T1]{fontenc}
\usepackage{ae,aecompl}

\usepackage{float}
\usepackage{graphicx}	
\usepackage{amsmath}	
\usepackage{amssymb}	

\title[Surface magnetism in R~Sct]{Surface magnetism in the pulsating RV Tauri star R Scuti\thanks{Based on observations obtained at the Télescope Bernard Lyot (TBL) at Observatoire du Pic du Midi, CNRS/INSU and Université de Toulouse, France.}}

\author[S. Georgiev et al.]{Stefan Georgiev,$^{1,2}$\thanks{E-mail: sgeorgiev@astro.bas.bg}
Agn{\`e}s L{\`e}bre,$^{2}$
Eric Josselin,$^{2}$
Philippe Mathias,$^{3}$\newauthor
Renada Konstantinova-Antova$^{1}$
and Laurence Sabin$^{4}$
\\
$^{1}$Institute of Astronomy and NAO, Bulgarian Academy of Science, 72, Tsarigradsko Chaussee Blvd., 1784 Sofia, Bulgaria\\
$^{2}$LUPM, Universit{\'e} de Montpellier, CNRS, place Eug{\`e}ne Bataillon, 34095 Montpellier, France\\
$^{3}$IRAP, Université de Toulouse, CNRS, UPS, CNES, 57 avenue d’Azereix, 65000 Tarbes, France\\
$^{4}$Instituto de Astronomia, Universidad Nacional Aut\'onoma de M\'exico, Apdo. Postal 877, C.P. 22860, Ensenada, B.C., M\'exico
}

\date{Accepted XXX. Received YYY; in original form ZZZ}

\pubyear{2022}

\begin{document}
\label{firstpage}
\pagerange{\pageref{firstpage}--\pageref{lastpage}}
\maketitle

\begin{abstract}
We present the surface magnetic field conditions of the brightest pulsating RV Tauri star, R~Sct. Our investigation is based on the longest spectropolarimetric survey ever performed on this variable star. The analysis of high resolution spectra and circular polarization data give sharp information on the dynamics of the atmosphere and the surface magnetism, respectively.  Our analysis shows that  surface magnetic field can be detected at different phases along a pulsating cycle, and that it may be related to the presence of a radiative shock wave periodically emerging out of the photosphere and propagating throughout the stellar atmosphere.
\end{abstract}




\section{Introduction}
\label{sec:rsct_intro}
\noindent RV~Tauri stars are high luminosity pulsating variables. Their lightcurves show alternating deep and shallow minima in a quasi-periodic manner, the periodicity being sometimes interrupted by irregular intervals due to intrinsic phenomena.
During regular parts of the lightcurve, the photometric variability period $P$ is defined as the time interval between two consecutive deep minima. The values of $P$ are usually between 30 and 150~d \citep{wallerstein84}.  \cite{Jura86} studied the mass loss rates of RV~Tauri stars and concluded that these objects have recently left an evolutionary phase of rapid mass loss. Considering their luminosities ($\sim 10^3L_{\odot}$) and formation rate (about 10\% of the rate of formation of planetary
nebulae) \cite{Jura86} proposed that they are post-AGB objects in transition from the Asymptotic Giant Branch (AGB) to the White Dwarf stage, which is a very short phase in the lifetime of a star.\\

\noindent R~Sct (HD~173819) is the brightest member of the RV~Tauri type stars. According to the General Catalogue of Variable Stars (GCVS\footnote{http://www.sai.msu.su/gcvs/gcvs/}), its visual magnitude ranges between 4.2 down to 8.6 and its photometric variability period is between 138.5 and 146.5~d (with variability from one cycle to another). Because of the strong photometric variability R~Sct exhibits, its spectral class also varies significantly: from G0Iae during maximum light to 
M3Ibe during deep minimum light. From observations taken during the phases of shallow minimum and just after a maximum in the lightcurve, \cite{kipper2013} measure the following stellar parameters: $T_{\rm eff} = 4500$~K, $\log g = 0.0$, [Fe/H]~=~-0.5.\\

\noindent For R\,Sct, during each pulsation cycle, two accelerations of the atmospheric layers occur, which are caused by the propagation of radiative shock waves \citep{gillet89}. By studying the variability of spectral lines, \cite{lebregillet91a, lebregillet91b} found that the expansion of the stellar atmosphere due to the passing of such shock waves may reach up to a few photospheric radii. Indeed, the propagation of these shocks produces significant ballistic motions in the atmosphere, that can be traced through dedicated spectral features, such as, for example, the splitting of metallic lines and emission in the lines of hydrogen. These studies also show that the two shockwaves associated to one pulsation cycle  (a main shock and a secondary shock) appear respectively just before the deep and the shallow light minimum, and that they may both be present at the same time, at different altitudes, in the higher part of the extended atmosphere of R~Sct. The authors also point out that this complex atmospheric dynamics may be at the origin of the mass loss the star experiences at this late evolutionary stage.\\

\noindent Since AGB and Post-AGB stars are among the main sources of enrichment of the interstellar medium through their winds and mass loss, the understanding of the origin of these processes is fundamental. An important mechanism, which would be related to mass loss, involves the presence of  magnetic fields. However, their exact role in this process at these late stages of stellar evolution is still poorly understood. For example, while binarity is  invoked as a principal shaping agent of the circumstellar envelope (CSE) \citep{Decin20}, magnetic fields could still play an active role in this task \citep{blackman2001}.\\

\noindent The first discovery of a surface magnetic field in R~Sct was reported by \cite{lebre2015}, who estimate the longitudinal magnetic field $B_l = 0.9 \pm 0.6$~G. Moreover \cite{sabin2015} found that the surface magnetic field of R~Sct is variable with time, sometimes being not present or below their detection limit of 0.5~G. They also observed a connection between the line profiles resulting from the shock propagation 
and the Zeeman signatures in circular polarization. On this basis they suggested there could be a link between the dynamical state of the extended stellar atmosphere of R~Sct and its surface magnetic field. Such a link has already been proposed for the pulsating variable Mira star $\chi$~Cyg \citep{lebre2014}, where the shockwave propagation throughout the stellar atmosphere seems to amplify, through compression, a 
weak surface magnetic field.\\

\noindent The objective of the present work is to investigate the magnetism of R~Sct during a long term spectropolarimetric survey, in order to check if there is indeed an interplay between the surface magnetic field and the atmospheric dynamics present in this cool evolved star. The article is organized as follows. Section~\ref{sec:rsct_obs_data_treatment} presents the spectropolarimetric observations used for the study, and the different tools used for their subsequent analysis. 
The atmospheric dynamics of R~Sct is then illustrated in Section~\ref{sec:rsct_dynamics} from the associated high resolution spectroscopic observations. In Section~\ref{sec:mag-field}, we present a refined approach to study the surface magnetism in such a pulsating star hosting  an extended atmosphere. The circular polarization data (Stokes~$V$ profiles) are analysed and the magnetic (Zeeman) origin of the detected signatures is investigated. 
Section~\ref{sec:rsct_linear} presents briefly the available observations of R~Sct in linear polarization, and finally, in Section~\ref{sec:rsct_summary}, a summary of the obtained results is given.

\section{Observations and data analysis}
\label{sec:rsct_obs_data_treatment}

\subsection{Spectropolarimetric observations}
\label{sec:rsct_obs}

\noindent Observations of R~Sct were obtained between July 2014 and August 2019 using the Narval instrument \citep{auriere2003} on the 2m class {\it Télescope Bernard-Lyot (TBL)} at the {\it Pic du Midi} observatory, France. This high resolution spectropolarimeter has a resolving power $R=65\,000$ and covers the spectral window between 375 and 1\,050 nm. The Narval instrument is designed to detect polarization levels (circular or linear) within atomic lines. It allows simultaneous measurement of the full intensity (Stokes~\textit{I}) and the intensity in circular (Stokes~\textit{V}) or linear (Stokes~\textit{U} or Stokes~\textit{Q}) polarization versus wavelength.\\

\noindent Our Narval observations represent the longest monitoring of R~Sct ever performed using high resolution spectropolarimetry so far. It corresponds to 6 seasons of observability of R~Sct obtained over the course of 6 years, which will hereafter be referred to as 6 datasets, from set 1 to set 6. All these Narval observations are composed of a number of spectropolarimetric sequences (circular and/or linear polarization),  all of which include a high resolution intensity spectrum (Stokes~\textit{I} measurements) and also a high resolution polarization spectrum (in Stokes~\textit{U}, \textit{Q} or \textit{V}).  With approximately one observation per month (during the seasons R~Sct is visible from {\it Pic du Midi}), linear and circular polarization data have been collected  preferably in a single night, or during consecutive nights depending on target observability or weather conditions. Hence, the total dataset spans over 67 different dates of observation, 31 of which contain Stokes~$V$ measurements, giving access to the information about the surface magnetism.  The full log of observations is given in Table~\ref{tab:log_observations}. The observations which contain measurements on linear polarization are presented in Section~\ref{sec:rsct_linear}, but the Stokes~$Q$ and $U$ profiles themselves are not examined in details in this study; instead, only their associated high resolution unpolarized spectra (Stokes~$I$) are used to trace the atmospheric conditions of R~Sct (in Section~\ref{sec:rsct_dynamics}). Our survey stops at the end of August 2019, when the TBL was stopped to install the successor of Narval, Neo-Narval. In the interests of keeping an homogeneous dataset, we therefore limited the data used in this study to the Narval observations only.

\begin{table*}
    \centering
    \begin{tabular}{lllll|lllll}
Date  &  HJD  &  Phase  &  Sequences  & SNR &  Date  &  HJD  &  Phase  &  Sequences & SNR\\
\hline
    2014/07/15  & 6854.5 & 0.57 &  1Q,1U,2V  & 994 &  2016/10/27  & 7689.3 & 6.47 &  1U & 868\\
    2014/07/22  & 6861.4 & 0.62 &  10V  & 765 &  2016/10/28  & 7690.3 & 6.48 &  2Q & 763\\
\cline{6-10}
    2014/09/01$^{\ddagger}$  & 6902.3 & 0.91 &  1Q,1U  & 1025 &  2017/04/18  & 7862.6 & 7.70 &  9V & 1179\\
    2014/09/02$^{\ddagger}$  & 6903.3 & 0.92 &  6V  & 828 &  2017/04/20  & 7864.6 & 7.71 &  2U & 1333\\
    2014/09/11$^{\ddagger}$  & 6912.3 & 0.98 &  1Q,6V,1U  & 855 &  2017/04/21  & 7865.6 & 7.72 &  2Q & 1340\\
\cline{1-5}
    2015/04/13  & 7126.6 & 2.50 &  3V  & 1021 &  2017/08/21$^{\ddagger}$  & 7987.4 &  --  &  6V & 1162\\
    2015/04/14  & 7127.6 & 2.50 &  1Q,1U  & 1067 &  2017/08/22$^{\ddagger}$  & 7988.4 &  --  &  2Q & 956\\
    2015/04/23  & 7136.6 & 2.57 &  2V  & 588 &  2017/09/02  & 7999.4 &  --  &  4Q & 753\\
    2015/05/27  & 7170.5 & 2.81 &  3V,1Q,1U  & 878 &  2017/09/06  & 8003.3 &  --  &  2Q & 1176\\
    2015/06/02  & 7176.6 & 2.85 &  9V  & 819 &  2017/09/07  & 8004.3 &  --  &  2U & 1650\\
    2015/06/26$^{\ddagger}$  & 7200.5 & 3.02 &  2U,2Q,9V  & 1447 &  2017/09/08  & 8005.3 &  --  &  3V & 1451\\
    2015/08/05  & 7240.4 & 3.30 &  2Q,2U  & 751 &  2017/10/02$^{\ddagger}$  & 8029.3 &  --  &  2U & 982\\
    2015/08/06  & 7241.9 & 3.31 &  16V  & 751 &  2017/10/04$^{\ddagger}$  & 8031.3 &  --  &  2Q & 541\\
    2015/08/10  & 7245.4 & 3.34 &  6V  & 1010 &  2017/10/05$^{\ddagger}$  & 8032.3 &  --  &  3V & 1051\\
    2015/08/26$^{\ddagger}$  & 7261.4 & 3.45 &  2Q  & 637 &  2017/10/30$^{\ddagger}$  & 8057.3 &  --  &  2U & 1338\\
    2015/08/28$^{\ddagger}$  & 7263.4 & 3.46 &  2U,6V  & 1234 &  2017/10/31$^{\ddagger}$  & 8058.2 &  --  &  2Q & 1121\\
    2015/10/08  & 7304.3 & 3.75 &  1Q,1U  & 796 &  2017/11/01$^{\ddagger}$  & 8059.8 &  --  &  9V & 1134\\
\cline{6-10}
    2015/10/09  & 7305.3 & 3.76 &  6V  & 796 &  2018/06/18  & 8288.5 & 0.05 &  2U,2Q & 1010\\
    2015/10/29  & 7325.3 & 3.90 &  1Q,1U  & 950 &  2018/06/19  & 8289.5 & 0.06 &  3V & 761\\
    2015/10/30  & 7326.3 & 3.91 &  5V  & 1414 &  2018/07/27  & 8327.5 & 0.32 &  2U,2Q & 868\\
\cline{1-5}
    2016/03/17$^{\ddagger}$  & 7465.6 & 4.89 &  1U,1Q  & 883 &  2018/07/28  & 8328.5 & 0.33 &  3V & 1142\\
    2016/05/04  & 7513.6 & 5.23 &  1U,1Q  & 1262 &  2018/08/15  & 8346.5 & 0.45 &  2Q,2U & 1166\\
    2016/05/15  & 7524.6 & 5.31 &  3V  & 1044 &  2018/08/31  & 8362.4 & 0.56 &  2Q,2U & 938\\
    2016/05/20  & 7529.6 & 5.34 &  3V  & 1105 &  2018/09/08  & 8370.4 & 0.61 &  3V & 1227\\
    2016/06/02  & 7542.6 & 5.44 &  1U,1Q  & 1083 &  2018/10/24$^{\ddagger}$  & 8416.3 & 0.92 &  2Q & 840\\
    2016/06/22  & 7562.5 & 5.58 &  1U,1Q  & 1141 &  2018/10/25$^{\ddagger}$  & 8417.3 & 0.93 &  2U & 954\\
\cline{6-10}
    2016/07/12  & 7582.5 & 5.72 &  1U,1Q  & 1436 &  2019/06/15  & 8650.5 & 2.51 &  3V,2Q & 1090\\
    2016/07/24  & 7594.4 & 5.80 &  3V  & 1147 &  2019/06/27  & 8662.5 & 2.59 &  3V & 1153\\
    2016/08/03  & 7604.4 & 5.87 &  1U,1Q  & 1049 &  2019/07/15  & 8680.4 & 2.71 &  2Q,2U & 1338\\
    2016/09/01  & 7633.3 & 6.08 &  1Q,1U  & 759 &  2019/07/18  & 8683.5 & 2.74 &  3V & 1274\\
    2016/09/24  & 7656.4 & 6.24 &  1Q  & 717 &  2019/08/05  & 8701.5 & 2.86 &  2U & 1088\\
    2016/09/28  & 7660.3 & 6.27 &  3V,1U  & 873 &  2019/08/13  & 8709.4 & 2.91 &  2Q & 847\\
    2016/10/03  & 7665.3 & 6.30 &  3V  & 1103 &  2019/08/31$^{\ddagger}$  & 8727.4 & 3.03 &  6V & 755
    \end{tabular}
    \caption{Log of spectropolarimetric observations. Calendar and heliocentric Julian dates (HJD) are given in the first two columns. HJD starts from 2~450~000. The horizontal lines show the limits of the 6 observational sets described in Section~\ref{sec:rsct_obs}. The phase (calculated according to Equations~1 or 2) is given in the third column. The fourth column indicates the number of Stokes~$Q$/$U$/$V$ sequences collected for each date. The fifth column shows the signal-to-noise ratio (SNR) of the Narval Stokes~$I$ spectra. The SNR is computed per spectral pixel at the center of the {\it echelle} order that contains the maximal signal (this typically corresponds to a wavelength of 731 or 755 nm). When multiple spectra are acquired on a given night, it is the median value of their SNR that is reported in the table. The dates for which the \textit{cool} mask is used in LSD are marked with $\ddagger$, and for the other dates, the standard mask is used (see Section~\ref{sec:rsct_masks_teff}).}
    \label{tab:log_observations}
\end{table*}

\begin{figure*} \centering
	\includegraphics[width = \linewidth]{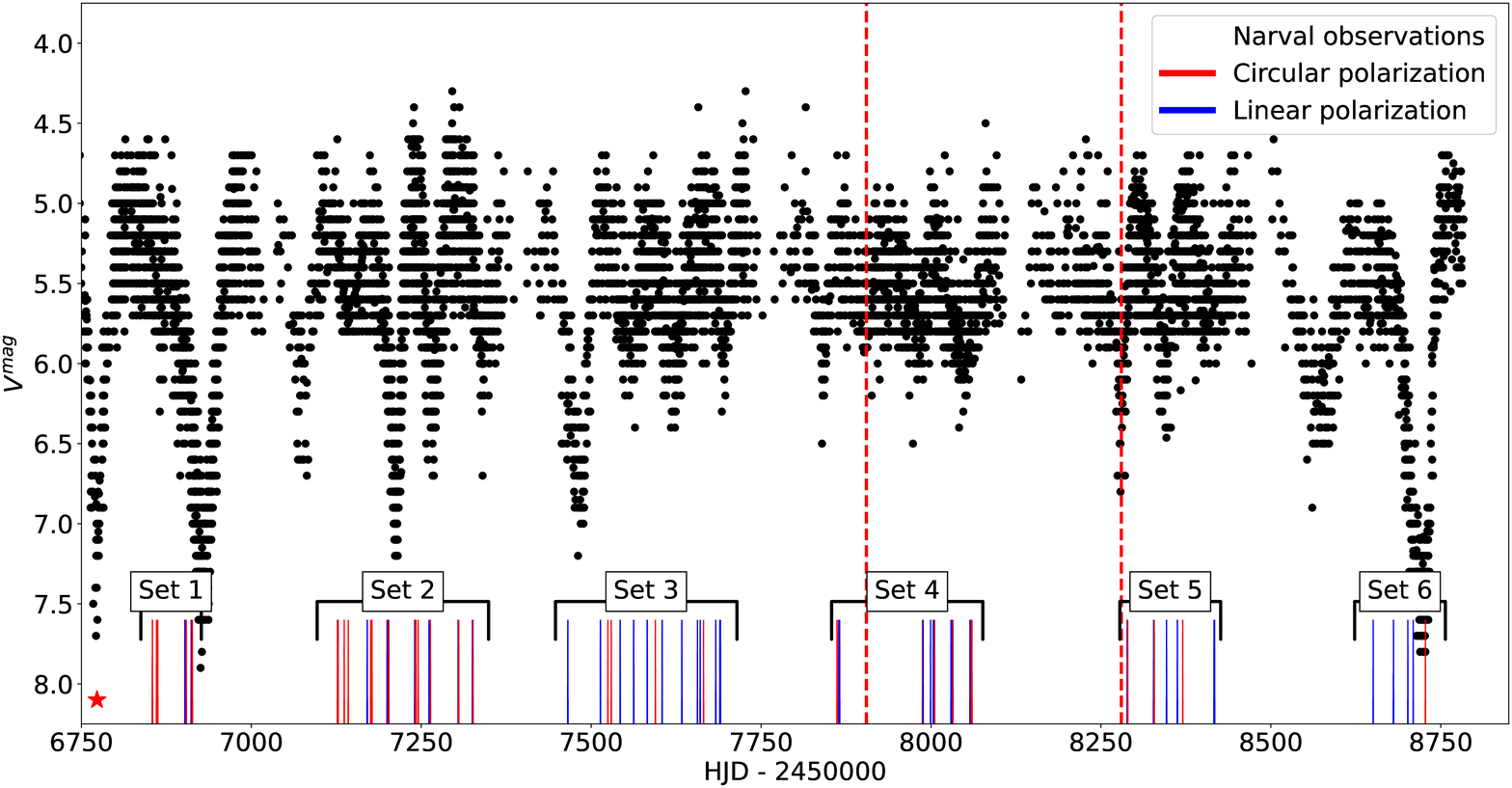}
	\caption[]{Visual lightcurve of R~Sct from AAVSO. The collected spectropolarimetric observations (see Table~\ref{tab:log_observations}) are indicated by vertical lines (red and blue lines for the circular and linear polarization observations, respectively) and the six observational sets are noted. The limits of the irregular interval of photometric variability (see Section~\ref{sec:rsct_obs}) are represented by the two dashed red  vertical  lines. The red star indicates the date selected to be the reference point (i.e. the beginning of period 0) in the first phase computation (see Equation~\ref{eq:rsct_eph1}); the reference point in the second phase computation (see Equation~\ref{eq:rsct_eph2}) corresponds to the rightest vertical dashed red line, i.e. the date the regularity of the lightcurve is restored.}
	\label{fig:rsct_lc}
\end{figure*}

\noindent In order to correlate the information on the atmospheric dynamics and the surface magnetism in our dataset to the photometric variability of R~Sct, the visual lightcurve of the star was retrieved from the AAVSO\footnote{\url{https://www.aavso.org/}} database, and it is presented in Figure~\ref{fig:rsct_lc}. The dates for which Narval observations  were collected are marked on the lightcurve with red and blue vertical lines for circular and linear polarization, respectively.\\

\noindent During the first part of the monitoring, between April 2014 (HJD = 2~456~773) and June 2016 (HJD = 2~457~905), the lightcurve of R~Sct appears regular (showing the usual alternation of deep and shallow minima). However, after this, the next expected deep minimum appears shallow instead, and after that the periodicity is lost. It is restored again in June 2019 (HJD 2~458~280). The limits of the irregular portion of the lightcurve between June 2016 and June 2019 are indicated on Figure~\ref{fig:rsct_lc} with dashed red vertical lines. This irregular portion within the light curve of R~Sct then only concerns most of the 4th dataset of our monitoring, while the five other datasets are related to regular portions of the light curve (see Fig.~\ref{fig:rsct_lc}). \\

\noindent The photometric period of R~Sct, when it is apparent (i.e. outside the irregular portion of the lightcurve in Fig.~\ref{fig:rsct_lc}) is variable from one cycle to the next one. We find that the average period length in the first regular part of the observational dataset is 141.50~d, while in the second one, after the irregular portion, it is 147.33~d. Using these two period values, and only for Narval observations associated to the regular parts of the lightcurve (see Fig.~\ref{fig:rsct_lc}) we compute the photometric phase according to the following ephemerids :

\begin{equation}
\label{eq:rsct_eph1}
    \phi(t) = \frac{t - 2456773.5}{141.5} \text{\; \; for sets 1, 2 and 3}
\end{equation}
\begin{equation}
\label{eq:rsct_eph2}
    \phi(t) = \frac{t - 2458280.5}{147.33} \text{\; \; for sets 5 and 6}
\end{equation}

\noindent We use Equation~\ref{eq:rsct_eph1} for the first regular portion of the lightcurve (in Figure~\ref{fig:rsct_lc}: between the red star located at HJD = 2\,456\,773.5 and the first vertical dashed red line located at HJD = 2\,457\,905.5) and Equation~\ref{eq:rsct_eph2} for the second one (in Figure~\ref{fig:rsct_lc}: after the second vertical dashed red line located at HJD = 2\,458\,280.5). We point out that the ephemerids are computed so that the deep minimum for each period corresponds to phase 0 (or to an integer value), which is the convention for RV~Tauri stars.

\subsection{Data Analysis}
\label{sec:rsct_masks_teff}

\subsubsection{Least Square Deconvolution (LSD)}
\label{sec:lsd}

\noindent In order to extract the mean polarization signatures from the data, we used the LSD method \citep{donati97} on our Narval data. This method is similiar to cross-correlation and uses a numerical line mask to average the profiles of thousands of atomic lines, under the assumption that they have the same shape scaled by a given factor. The result of the LSD procedure is a mean line profile, referenced later as the LSD profile, in heliocentric velocity space, in both full intensity (Stokes~\textit{I}) and polarized light (Stokes~\textit{Q/U/V}).\\

\noindent For a given date, when several spectropolarimetric sequences are available (see Table \ref{tab:log_observations}), we first performed LSD on the individual sequences, and then combined the obtained results into an averaged LSD output (displayed and analysed through this paper). We note that occasionally for some observations the individual sequences were obtained in two consecutive nights. 
In these cases we average together the LSD outputs of the sequences from consecutive nights, i.e. we consider them as a single observation. Such a combining of sequences from consecutive nights is necessary to achieve higher signal-to-noise ratio when using LSD. This is possible in the case of R~Sct because there is no noticeable variability on such short time-scales (1 day) between the individual sequences (neither in the intensity spectra, nor in the polarized profiles). \\

\begin{figure} \centering
    \includegraphics[width = \linewidth]{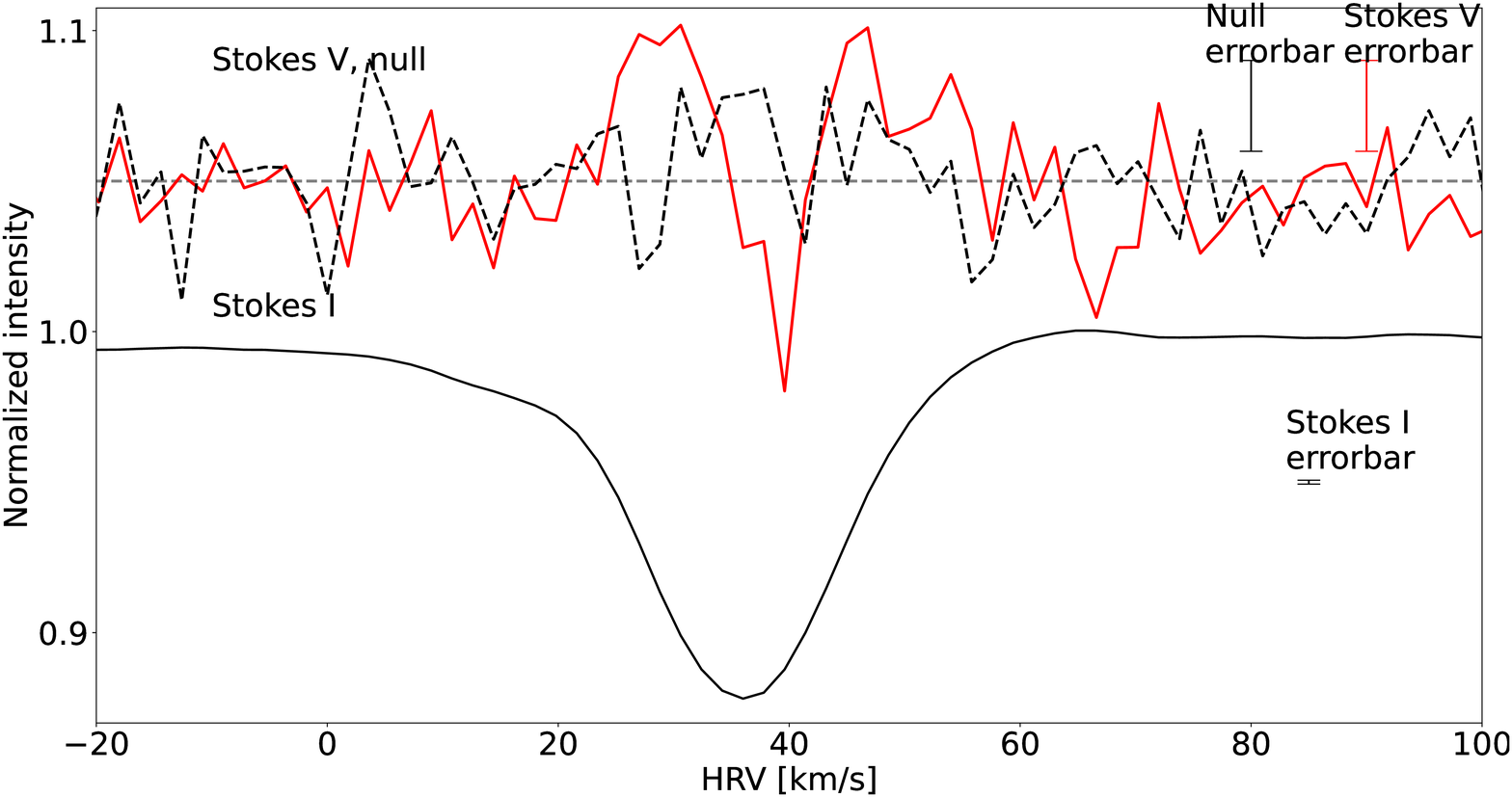}
	\caption[]{Example of the LSD profiles (Stokes~\textit{I}, solid black curve, and  Stokes~\textit{V}, solid red curve) for R~Sct at the date of 2014/09/02, corresponding to a definite detection (see Section~\ref{sec:detection}). The dashed black curve displays the null diagnostic, that has to stay featureless to confirm the absence of instrumental effect. The null and Stokes~\textit{V} profiles are shifted vertically and magnified for display purposes, and their zero level is indicated with a dashed horizontal line. The mean errorbars of the Stokes~$I$, null and Stokes~$V$ profiles are also shown and labelled.}
	\label{fig:example_lsd_figure}
\end{figure}

\noindent To study the surface magnetic field in R~Sct, we applied the LSD method on the  observational data to extract the mean Stokes~\textit{I} and  Stokes~\textit{U/Q/V} profiles of atomic spectral lines.
In order to do this, we first built a line mask computed from a MARCS model atmosphere \citep{gustafsson2008} with parameters compatible with the known parameters of the star ($T_{\rm eff}~=~4500$~K, $\log g~=~0$, microturbulence velocity $\xi~=~2$~km/s and [Fe/H]~=~-0.5), and using atomic linelists extracted from the Vienna Atomic Line Database, VALD \citep{kupka99}. We kept the atomic lines of all chemical elements except those elements that have lines known to exhibit emission and/or trace the very high atmosphere (H, He, Ca) and lines known to be affected by interstellar contamination (e.g., Na and K). Only lines with a central depth greater than 0.2 of the level of the continuum were considered in the mask computation.
An example of an LSD output (computed with such a numerical mask gathering about 17\,000 atomic lines) is shown in Figure \ref{fig:example_lsd_figure}.\\

\noindent An LSD profile is normalized by three independent coefficients. These are the equivalent wavelength, equivalent depth and equivalent Land{\'e} factor ($\lambda_{\rm eff}$, $d_{\rm eff}$ and $g_{\rm eff}$, respectively). These three coefficients do not, in general, correspond to the respective mean values of the wavelength, depth and Land{\'e} factor of the atomic lines used in the LSD procedure. In order to avoid differences in this normalization when comparing LSD profiles of different observations (as it is done in this work), in \cite{donati97} two parameters are introduced that are functions of $\lambda_{\rm eff}$, $d_{\rm eff}$ and $g_{\rm eff}$: a mean intensity weight $W_{\rm int}$ and a mean polarization weight $W_{\rm pol}$. When both of these weights are equal to 1, the resulting LSD profile does not depend on $\lambda_{\rm eff}$, $d_{\rm eff}$ and $g_{\rm eff}$. In this paper, all LSD calculations are performed in this way, with $W_{\rm int} = W_{\rm pol} = 1$, which allows the comparison of LSD profiles obtained on different nights of observation.\\

\subsubsection{Atomic line masks}
\label{sec:lsd_masks}

\noindent In the course of its photometric variability, R~Sct changes significantly not only its visual magnitude, but also its spectral type, hence its effective temperature $T_{\rm eff}$. From the literature it is known that during a shallow  minimum light and around a maximum luminosity, the effective temperature of the star is 4500~K \citep{kipper2013}. However the spectral type changes to a much later one near a deep minimum light, suggesting different atmospheric conditions through a lower effective temperature. Our Narval observations also show that TiO molecular bands appear in the  high resolution spectra of R~Sct collected close to a deep minimum light, indicating a significant decrease in $T_{\rm eff}$. This means that using a single LSD mask computed with a unique effective temperature reference to treat our whole dataset may not be relevant. A more reliable result would be obtained if, instead, the observations obtained around a deep minimum light (hence showing these TiO band features) are treated using a LSD mask computed for a lower effective temperature closer to that of the star at this photometric stage.\\

\noindent To estimate the effective temperature of R~Sct around a deep minimum light, we used the code \textit{Turbospectrum}  \citep{plez, alvarez_plez98} to construct a series of synthetic spectra around two molecular bands of TiO (TiO~7054 {\AA} and TiO~7589 {\AA}) which are detected in the spectra of R~Sct during these specific phases.
To construct the synthetic spectra, we used again linelists extracted from the VALD database and MARCS model atmospheres computed for $1M_{\odot}$, $\log g = 0.0$, $\xi~=~2$~km/s and $T_{\rm eff}$ = 4500, 3800, 3600 and 3400~K. We used the built-in procedure in \textit{Turbospectrum} to take into account the instrumental profile of Narval and to introduce the effects of rotational and macroturbulent velocities. Regarding the latter two velocities, neither value is known for R~Sct, and we assigned to $v \sin i$ and to $v_{\rm mac}$ the \textit{ad hoc} value of 5 km~s$^{-1}$, which is a reasonable value for AGB and post-AGB stars \citep{Georgiev20}. It must be noted that there is no particular reason to pick these exact values, and that they have no impact on the task at hand. The objective is indeed not to determine accurate stellar parameters, but instead only to give a rough estimation of $T_{\rm eff}$; for this purpose, any reasonable pair of these two velocities would suffice.\\

\begin{figure}
	\includegraphics[width = \linewidth]{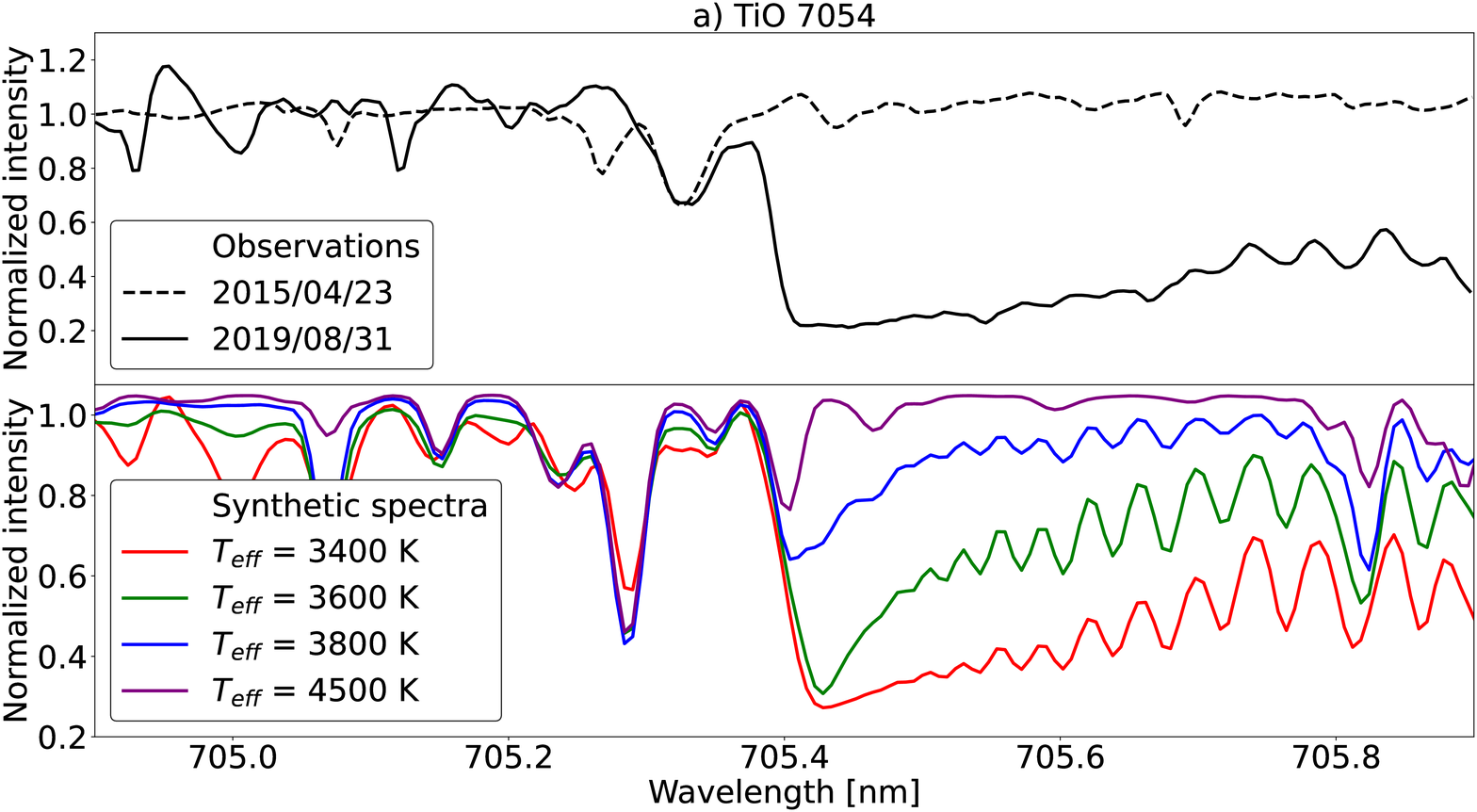}
	\includegraphics[width = \linewidth]{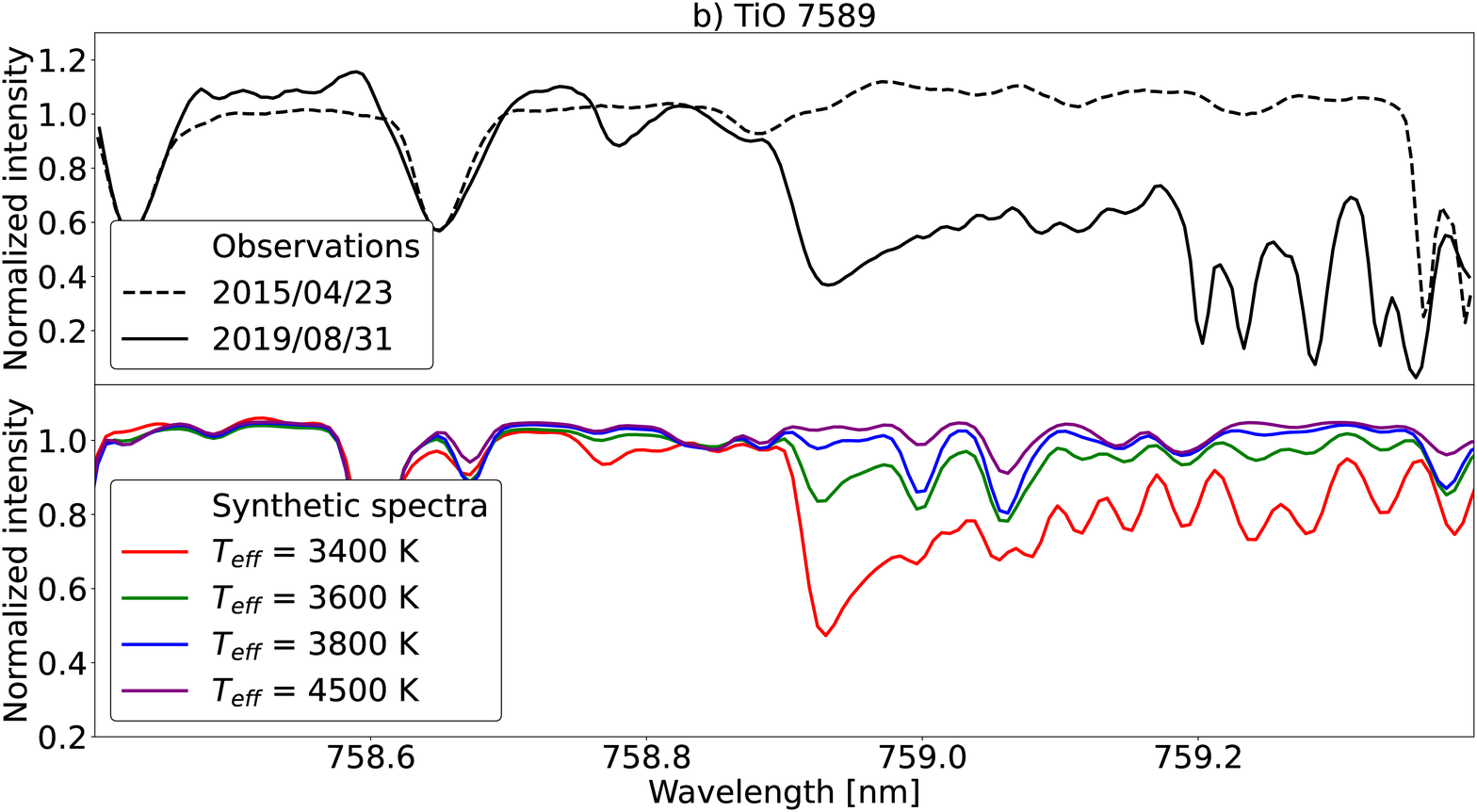}
    \caption[]{TiO 7054 {\AA} (panel "a") and TiO 7589 {\AA} (panel "b") bands in the spectra of R~Sct. For each TiO band, the top subpanel shows two observed spectra, obtained on 2015/04/23 (corresponding to a shallow  minimum) and 2019/08/31 (corresponding to a deep  minimum). The bottom subpanels show synthetic spectra computed for four different values of $T_{\rm eff}$.}
	\label{fig:rsct_teff_tests}
\end{figure}

\noindent In Figure~\ref{fig:rsct_teff_tests} we compared the computed synthetic spectra to two observational spectra of R~Sct, obtained on 2015/04/23 (corresponding to a shallow minimum light) and on 2019/08/31 (collected at the exact time of a deep minimum light). It can be seen from the synthetic spectra that the 7054 {\AA} TiO band (Figure~\ref{fig:rsct_teff_tests}a) begins to develop only at temperatures below 3800~K. On the other hand, this band is present in the observation collected around a deep minimum light (solid black line), but not in the one collected around a shallow  minimum light (dashed black line). This is consistent with the 4500~K estimation \cite{kipper2013} obtain for the shallow minimum and maximum phases. The synthetic spectra also show that this TiO band continues to develop in both depth and width as temperature decreases until $T_{\rm eff} = 3400$~K. Further decreasing the temperature makes the band more shallow instead. As displayed in Figure~\ref{fig:rsct_teff_tests}b, the 7589 {\AA} TiO band) is again only present in observations obtained around a deep minimum light (solid black line).
In the synthetic spectra, this band is not present at 4500~K and not even at 3800~K. It begins to develop only at 3600~K and grows in depth and width as temperature decreases, reaching noticeable depth in the 3400~K spectrum. We can draw two firm conclusions from this: 1) around a shallow  minimum light, the effective temperature of R~Sct is above 3800~K, in good agreement with the measurements of \cite{kipper2013}; 2) around a deep minimum light, the effective temperature is below 3800~K, since both TiO bands appearing in the spectra are best described by the synthetic spectra computed for 3400~K and 3600~K (see Fig.~\ref{fig:rsct_teff_tests}), which well matches the M spectral type expected during times of deep minimum.\\

\noindent For this reason, we decided to perform LSD in the following manner:\\
\indent{-- for all observations in which TiO bands are absent, i.e. all except those obtained around a deep minimum light - using a mask computed for $T_{\rm eff} = 4500$~K (henceforth, the \textit{standard} mask);}\\
\indent{-- for the only observations in which TiO bands are present, i.e. those obtained around a deep minimum light - using a mask computed for $T_{\rm eff} = 3500$~K (henceforth, the \textit{cool} mask).}\\

\noindent A last justification for the use of such a \textit{cool} mask (at deep minimum phases) is that even if it does not perfectly reflect the effective temperature of R~Sct during a deep minimum light, it certainly gives a better approximation than the \textit{standard} mask does.
 The other stellar parameters, the \textit{standard} and \textit{cool} masks were computed for, are $\log g = 0.0$, microturbulence of 2~km~s$^{-1}$ and [Fe/H]~=~-0.5, in agreement with \cite{kipper2013}. The total number of lines in each of the two masks is approximately 17~000, with the \textit{standard} mask containing slightly more strong lines (with a central depth larger than the average) than the \textit{cool} one.\\

\begin{figure}
    \centering
	\includegraphics[width = \linewidth]{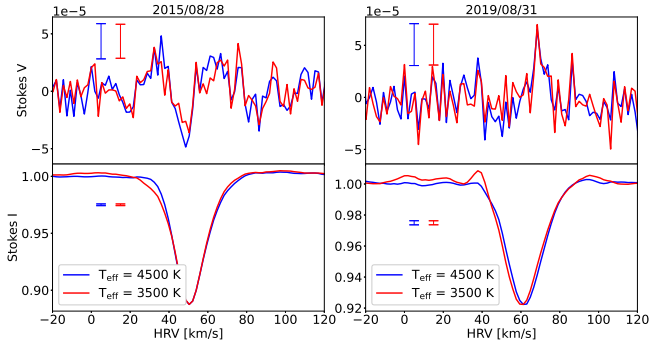}
	\caption[]{Comparison between the LSD profiles obtained with the \textit{standard} (blue) and \textit{cool} (red) masks for two observations obtained around a deep minimum light -- 2015/08/28 (left) and 2019/08/31 (right). The upper panels show the respective Stokes~$V$ profiles, while the bottom panels show the intensity profiles for both observations. The mean errorbars of the Stokes~$I$ and Stokes~$V$ profiles are indicated for both masks.}
	\label{fig:rsct_hot_cool_mask_comparison}
\end{figure}

 \noindent In Figure~\ref{fig:rsct_hot_cool_mask_comparison}, a comparison between the LSD profiles obtained with the \textit{standard} (in blue) and \textit{cool} (in red) masks is presented for two observations obtained around a deep minimum light (2015/08/28 and 2019/08/31). It can be seen that differences are present in the shape and amplitude of both the Stokes~$V$ (upper panels) and the intensity profiles (bottom panels), which justifies the use of the \textit{cool} mask for observations where TiO bands are present in the spectrum.
 
 \noindent As we shall see in Section~\ref{sec:mag-field}, the selection of lines in these LSD masks can be constrained even further in order to trace only the contribution of the photosphere (in this work for convenience by "photosphere" we mean the bottom part of the atmosphere).
  But first, in the following section, we infer the atmospheric dynamics of R~Sct along our Narval monitoring (covering about 2\,000 days), by studying spectral indicators. 

\subsubsection{Detection of polarized signatures}
\label{sec:detection}

\noindent In order to evaluate the presence or lack of signature in the polarized LSD profile, we used the \textit{specpolFlow} package\footnote{\url{https://github.com/folsomcp/specpolFlow}} for treatment of spectropolarimetric observations. This package provides a routine for the calculation of the probability that the observed polarized signal is consistent with the null hypothesis of no magnetic field, or the false alarm probability (FAP), evaluated from $\chi^2$. We then take the calculated value of the FAP and apply to it the criteria defined by \cite{donati97} in order to classify an observation as a no detection (ND, FAP~$>~10^{-2}$), a marginal detection (MD, $10^{-2}~>$~FAP~$>~10^{-4}$), or a definite detection (DD, FAP~$<~10^{-4}$). This detection parameter is provided in column 4 of Table~\ref{tab:log_stokesv}.

\section{Atmospheric dynamics}
\label{sec:rsct_dynamics}

\noindent The first qualitative model of the pulsation of the photosphere of R~Sct was introduced by \cite{gillet90}. From spectroscopic observations, these authors reported that two shockwaves propagate throughout the stellar atmosphere along one pulsation cycle.
The first (main) shockwave is found to emerge from the photosphere just before the phase of deep photometric minimum (set at $\phi = 0$). This main shock is strong and it is  supposed to produce a large elevation of the atmosphere, followed by a subsequent infalling motion due to the force of gravity (ballistic motion). A secondary (and weaker) shockwave emerges from the photosphere just before the occurrence of a shallow photometric minimum around $\phi = 0.5$, while the main shock is vanishing in the upper atmosphere.\\

\noindent This work was extended by \cite{lebregillet91a, lebregillet91b} who investigated the profile variations of selected spectral lines which turned out to be good indicators of the motions at the photospheric level, and of ballistic motions taking place in the upper atmosphere. 

\subsection{Variability of the H$\alpha$ line}
\label{sec:rsct_ha}

\begin{figure*}
	\includegraphics[width = 0.33\linewidth]{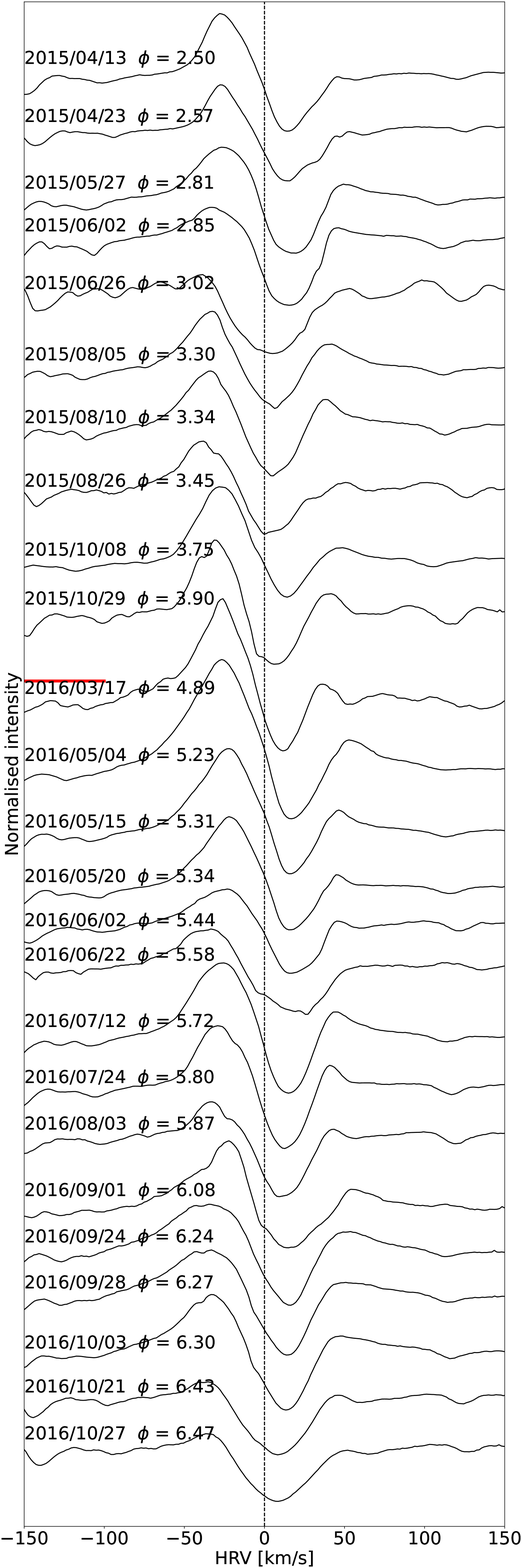}
	\includegraphics[width = 0.33\linewidth]{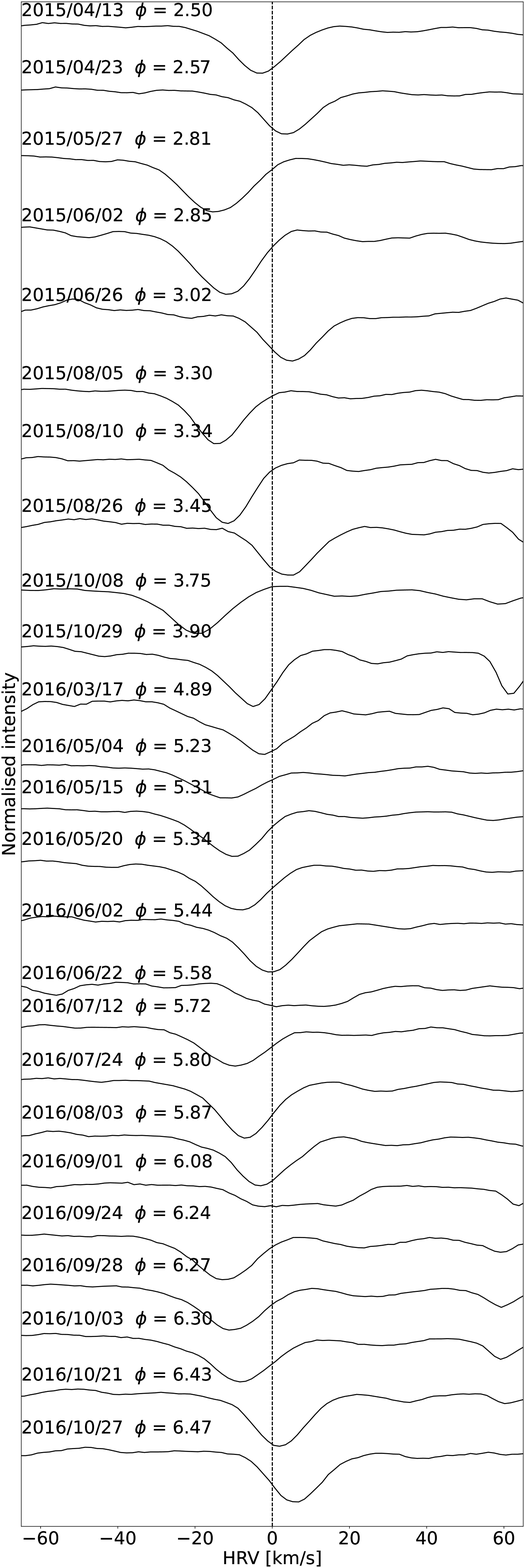}
	\includegraphics[width = 0.33\linewidth]{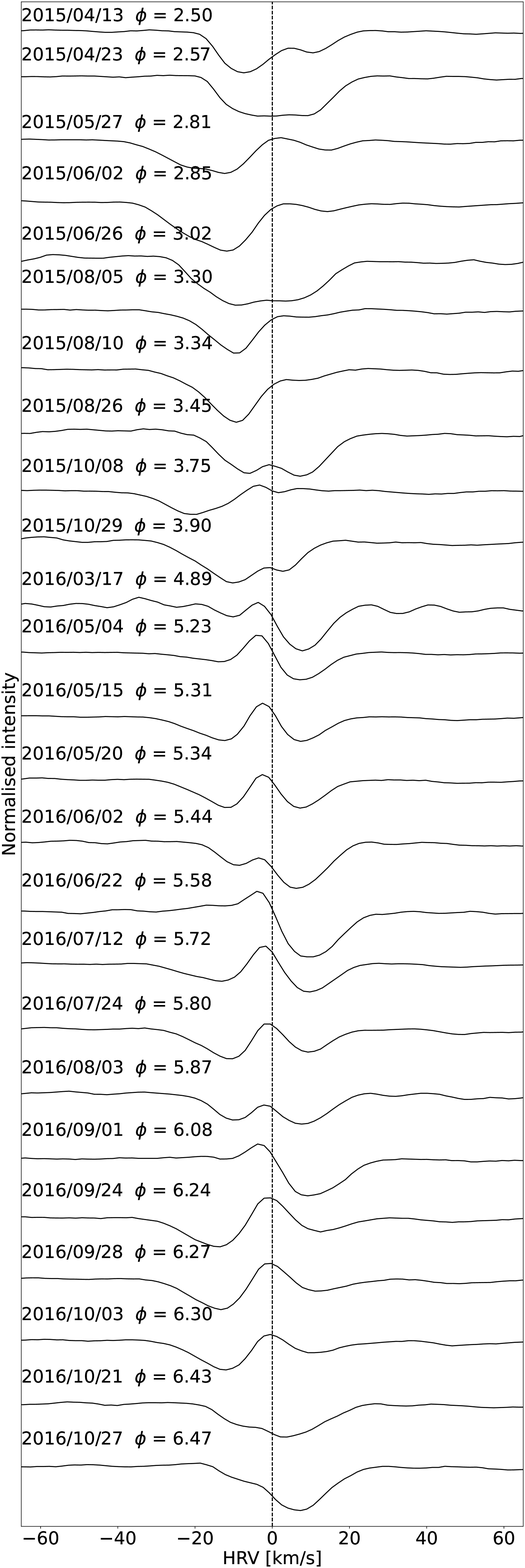}
	\caption[]{Time variation of the H$\alpha$ line (left panel), FeI~6569~${\AA}$ (middle) and TiI~5866~${\AA}$ (right) profiles in the spectra of R~Sct during datasets 2 and 3. The red horizontal line indicates the boundary between set 2 and set 3. Each observation is shifted vertically for display purposes, and its date is indicated on the top left side, with its phase (computed from Equation~\ref{eq:rsct_eph1}). The restframe velocity of R~Sct is indicated with a vertical dashed line. When observations from consecutive dates are available, only one date is presented, since no visible difference in the line profile can be appreciated on such a very short time-scale.}
	\label{fig:rsct_ha_fei_tii}
\end{figure*}

\noindent In our Narval spectra of R~Sct, the H$\alpha$ line shows an emission component which varies in strength along the photometric phase. The left panel of Figure~\ref{fig:rsct_ha_fei_tii} shows the profile variation of the H$\alpha$ line along the  observational datasets 2 and 3.  We selected these two sets because they offer the best phase coverage with respectively 15 and 16 different observational dates  (representing $\sim 50\%$ of the total number of observations in our survey of R~Sct).
Hence the time variations of dedicated line profiles over several pulsation cycles covering a large portion of the regular part of the lightcurve (see Fig.~\ref{fig:rsct_lc} and Tab.~\ref{tab:log_observations}) can be investigated. In our study, we use the value of the stellar restframe velocity $v_{\rm rad}$ = 42.47~km/s, recently derived by \cite{kunder2017}. We shall use this value only as a reference point to label the components of spectral line profiles either "blueshifted" or "redshifted" with respect to it.\\

\noindent It can be seen in Fig.~\ref{fig:rsct_ha_fei_tii} that H$\alpha$ shows an emission component with variable amplitude, which is blueshifted with respect to the restframe velocity ($v_{\rm rad}$) of R~Sct. H$\alpha$ also shows a variable profile which consists of an absorption component, usually redshifted with respect to $v_{\rm rad}$. It can be also seen that the emergence of the main shockwave from the photosphere (expected around the phase $\phi = 0$) usually results into a strong H$\alpha$ emission component (e.g. on 2015/10/29, before the  minimum light occurring on 2015/11/13 and on 2016/03/17, before the  minimum light occurring on 2016/04/02). This is in agreement with the work of \cite{gillet89}, who presented the complex H$\alpha$ profile as a full emission originating from the front of a radative shock wave, i.e., from a narrow region where the temperature and pressure conditions are favorable to produce such an intense emission line; then the cool hydrogen (either static or relaxing) located above the shock front, introduces an absorption component to the central and/or redshifted part of the line. The secondary shock, while weaker than the main one, is also detectable through an H$\alpha$ emission around the phase $\phi = 0.5$ (i.e., around a shallow photometric minimum). This emission feature then decreases until a new intense emission occurs due to the emergence of a new main and strong shock. A few dates also show a very faint H$\alpha$ emission (e.g., 2015/06/26 at $\phi = 3.02$) or even no emission at all (e.g., 2016/10/27 at $\phi = 6.47$) illustrating the few moments when the atmosphere is almost homogeneously relaxing in a global infalling motion.

\subsection{Variability of metallic lines}
\label{sec:rsct_metallic_lines}

\noindent To study the motions of the atmospheric layers of R~Sct during the longest high resolution spectropolarimetric monitoring ever realized for a post-AGB star, we explore the profile variations of metallic lines. \cite{schwarzschild52} proposed a mechanism to explain the  variations of atomic spectral lines detected in variable pulsating stars hosting shock waves propagating throughout their atmosphere. 
Indeed the passing of a shock wave through the stellar atmosphere first makes the layers closest to the photosphere move upwards (towards the observer), thus causing the spectral lines formed in these layers to be blueshifted; after the shock has passed, these same layers, as they are relaxing, fall ballistically downwards (away from the observer),  thus causing the spectral lines to be redshifted. This  Schwarzschild's mechanism also predicts the possibility for some metallic lines to appear through a double profile, when the shock is strong enough to decouple two velocity regions along the same line of sight. In that case, the bottom part of a large layer of line formation is moving upwards (as lifted up by the passage of a shock wave) while its upper part, located above this shock front, is still relaxing.
 This basic scenario may be even more complicated in the case of a very extended atmosphere, where the imprints -on atomic metallic lines- of two consecutive shockwaves may be present at the same time, as it has been shown to be the case for R~Sct by \cite{lebregillet91b}. \\

\noindent In the spectra of R~Sct, atomic lines of elements such as Fe and Ti  display profiles that vary with the photometric phase, being either fully blueshifted, double-peaked or fully redshifted. 
According  to \cite{lebregillet91a, lebregillet91b}, 
profile variations of FeI line (${\lambda}\,6569$ \AA) and TiI line (${\lambda}\,5866$ \AA) are  good  diagnostics to investigate the pulsation motions of their formation regions: the photospheric region, and the upper atmosphere, respectively. Those lines appear to be free from atomic or molecular blend, even at the coolest phases along the pulsation cycle. 

\subsubsection{The lower atmosphere of R~Sct}
\label{sec:rsct_fei}

\noindent The profile variation of the FeI 6569 {\AA} line during observational sets 2 and 3 (between April 2015 and October 2016, see Table~\ref{tab:log_observations}) is shown in the middle panel of Figure~\ref{fig:rsct_ha_fei_tii}. It can be seen from this figure that this line mostly shows a single gaussian profile (with the exception of very few dates such as 2016/06/22 and 2016/09/01), either blueshifted or redshifted with respect to the restframe velocity of R~Sct. Following the Schwarzschild mechanism  \citep{schwarzschild52}, this corresponds respectively to phases  at  which the layer of formation of this atomic line is rising (for the blueshifted profiles) or falling (for the redshifted ones). The absence of a clear double-peak profile  of the FeI 6569 {\AA} line in most of the observations shows that the layer in which this line forms is very narrow and thus is affected only by a single global motion (upward or downward) at any given time.  \cite{lebregillet91a} showed that the FeI 6569 {\AA} line is formed in a narrow region close to the photosphere, this line can thus be used to trace the motion of the photosphere, and hence the periodic emergence of shockwaves.\\

\begin{figure} \centering
	\includegraphics[width = \linewidth]{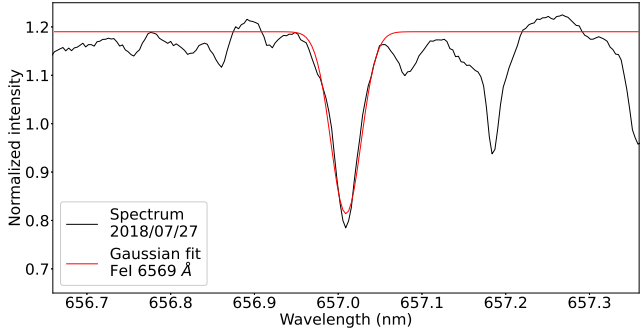}
	\caption[]{An example of a gaussian fit (in red) to the FeI 6569 {\AA} line in the spectrum of R~Sct obtained on 2018/07/27 (in black).}
	\label{fig:example_fei_fit}
\end{figure}

\begin{figure} \centering
	\includegraphics[width = \linewidth]{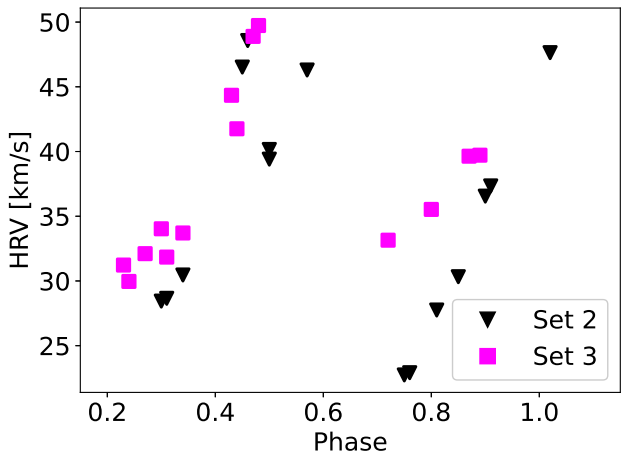}
	\caption[]{Radial velocity curve of the near-photospheric layers of the atmosphere of R~Sct obtained from gaussian fits on the FeI 6569 {\AA} line. Points obtained from set 2 and from set 3 are displayed with different colored symbols.}
	\label{fig:rsct_fei_rv}
\end{figure}

\noindent In order to trace the emergence of shockwaves through the photosphere of R~Sct, the radial velocity curve of the FeI 6569 {\AA} line was constructed by fitting its profile with a gaussian function in the spectra of R~Sct obtained during observational sets 2 and 3. An example fit is shown in Figure~\ref{fig:example_fei_fit} for the observation taken on 2018/07/27. The radial velocity curve is shown in Figure~\ref{fig:rsct_fei_rv}. In this figure, a regular global motion (up and down) of the explored photospheric region  is clearly seen. It can  be noticed that  there are indeed two upward accelerations per photometric period, found around phases $\phi = 0$ and $\phi = 0.5$, which correspond respectively to the times of deep and shallow photometric minima, confirming the results of \cite{gillet89}. The observed dispersion in the radial velocity curve most likely corresponds to cycle-to-cycle variations, as the intensity of the shocks is probably different for one cycle to the other.\\

\subsubsection{The upper atmosphere of R~Sct}
\label{sec:rsct_tii}

\noindent \cite{lebregillet91a}  showed that the TiI 5866 {\AA} line is formed in the upper atmosphere of R~Sct, as this line appears more perturbed by ballistic motions along a pulsation cycle. The time variability of the profile of this spectral line during sets 2 and 3 is shown in the right panel of Figure~\ref{fig:rsct_ha_fei_tii}. We see that the behavior of this line profile is indeed very different from that of the FeI 6569 {\AA} one: complex and strongly variable profiles, both single and double-peaked, are observed on different dates. Following the Schwarzschild mechanism, since for some phases we can report  double-peaked profiles (e.g. on   2015/08/26,   2016/05/15, 2017/10/05),  this suggests that the TiI 5866 {\AA} line is formed in a large geometric layer of the upper  atmosphere that can be affected -  at a given date - by more than one dynamics (i.e, upward and downward motions within the considered thick layer). At the dates double line profiles are reported, at least two different dynamics are thus present within this extended atmospheric layer, resulting from the outward propagation of a shockwave. On the contrary very few dates (e.g. on 2015/08/05, 2016/03/17) present  the TiI 5866 {\AA} line through a single gaussian profile (blueshifted and redshifted, respectively) suggesting the atmospheric layer is affected by a global motion. 

\subsection{Atmospheric dynamics: summary}
\label{sec:rsct_atm_summary}

\noindent The analysis of the variability of the H$\alpha$ and FeI 6569 {\AA} line profiles (however limited to data from set 2 and set 3) confirms the results of \cite{gillet89}, who found that two shockwaves per photometric period emerge from the photosphere of R~Sct around the deep and shallow photometric light minima (or,
equivalently, just before phases $\phi = 0$ and $\phi = 0.5$). Considering also the TiI 5866 {\AA} line variability, it is obvious that spectral lines originating from layers of different altitudes and different thickness are also affected by the propagating shockwaves in different ways: "photospheric" lines (i.e. lines which form in a very narrow region near the photosphere, such as the FeI 6569 {\AA} line) always point to a single global motion -- upward or downward -- when affected by the passing of a new shockwave, and also indicate the moment at which such a shock emerges from the photosphere; lines which form in wider regions of the atmosphere (such as the TiI 5866 {\AA} line), on the oher hand, may trace more than one dynamics, depending on the location of the shockwave.

\section{The surface magnetic field}
\label{sec:mag-field}

\subsection{A refined approach}
\label{sec:rsct_refined_lsd}

\noindent The LSD Stokes~$V$ profiles computation described in Section~\ref{sec:rsct_masks_teff} is undertaken along the "classical" way  when studying surface magnetic fields using high resolution spectropolarimetry. However, before going any further 
in our analysis, we must consider that this usual approach may not be the best one to use in the case of stars hosting strong pulsations and shockwaves propagating in their atmospheres, such as R~Sct. These stars have indeed very extended atmospheres and,  at a given time, different atmospheric layers might be affected by very different dynamics. This is precisely what has been illustrated through the investigation of the FeI 6569 {\AA} and TiI 5866 {\AA}  lines in  Section~\ref{sec:lsd_masks}. Thus, an LSD line mask computed without considering the conditions under which spectral lines form (like the \textit{standard} and \textit{cool} masks involved in Section~\ref{sec:lsd_masks}) will surely mix together information from very different altitudes with various physical conditions, surely affected, at any given moment in time, by different dynamics. This means that the use of line masks described in Section~\ref{sec:lsd_masks} would yield LSD profiles combining the full complexity of the atmospheric motions, which may however be different for the photospheric region and for the upper atmosphere. Because our objective is to study the \textit{surface} magnetic field (i.e. its condition at the level of the photosphere), introducing information from the higher atmosphere is certainly not the best approach. Hence, new LSD masks have to be constructed  based on a refined method for lines selection.

\subsection{Lines selection by excitation potential}
\label{sec:rsct_line_selection_chi}

\noindent The higher the excitation potential ($\chi$) of atomic lines is, the less extended their zone of formation in the stellar atmosphere is. In particular, lines with large $\chi$ can only be formed in a thin layer near the photosphere. This narrow region is expected to be affected, at any given moment in time, by only one dynamical effect: an upward global motion due to the presence of a  shock wave passing through it, or a downward global motion due to the relaxation of this  layer after the passage of such a shock. For example, the FeI 6569 {\AA} line, which was shown in Section~\ref{sec:rsct_fei} to be formed only in a narrow region near the photosphere of R~Sct, has an excitation potential of 4.7~eV (according to the VALD data). On the other hand, lines which have a low excitation potential can be formed under a wider variety of conditions, met both at the photospheric layers and also at higher altitudes. These lines are thus expected to combine the complexity of a large portion of the stellar atmosphere, which may contain layers affected by different dynamics -- upward and downward motions. For example, the TiI 5866 {\AA}, shown in Section~\ref{sec:rsct_tii} to be formed in an extended region of the atmosphere of R~Sct, has an excitation potential of 1.06~eV (according to the VALD data).\\

\noindent This effect is observed for all lines with similar excitation potentials, but in the case of cool stars, it can be illustrated for only a handful of metallic lines, which are not affected by molecular veiling, or through a cross-correlation profile, as done by \cite{alvarez2000} for Mira stars and by \cite{josselinplez2007} for red supergiant stars.\\

\begin{figure}
    \centering
	\includegraphics[width = 0.725\linewidth]{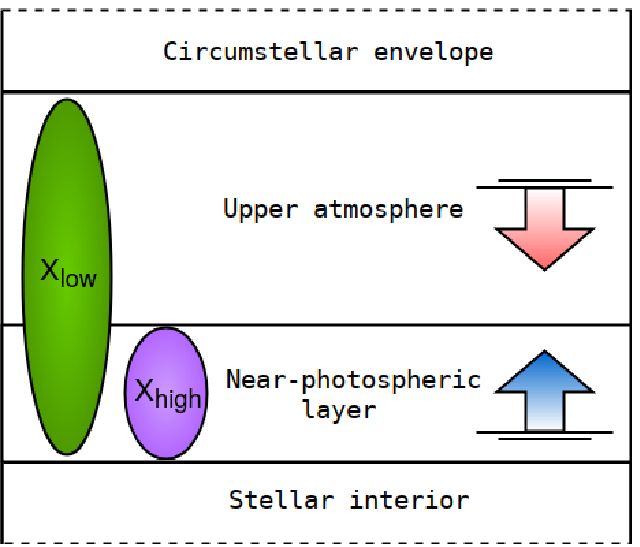}
	\centering
   	\includegraphics[width = 0.755\linewidth]{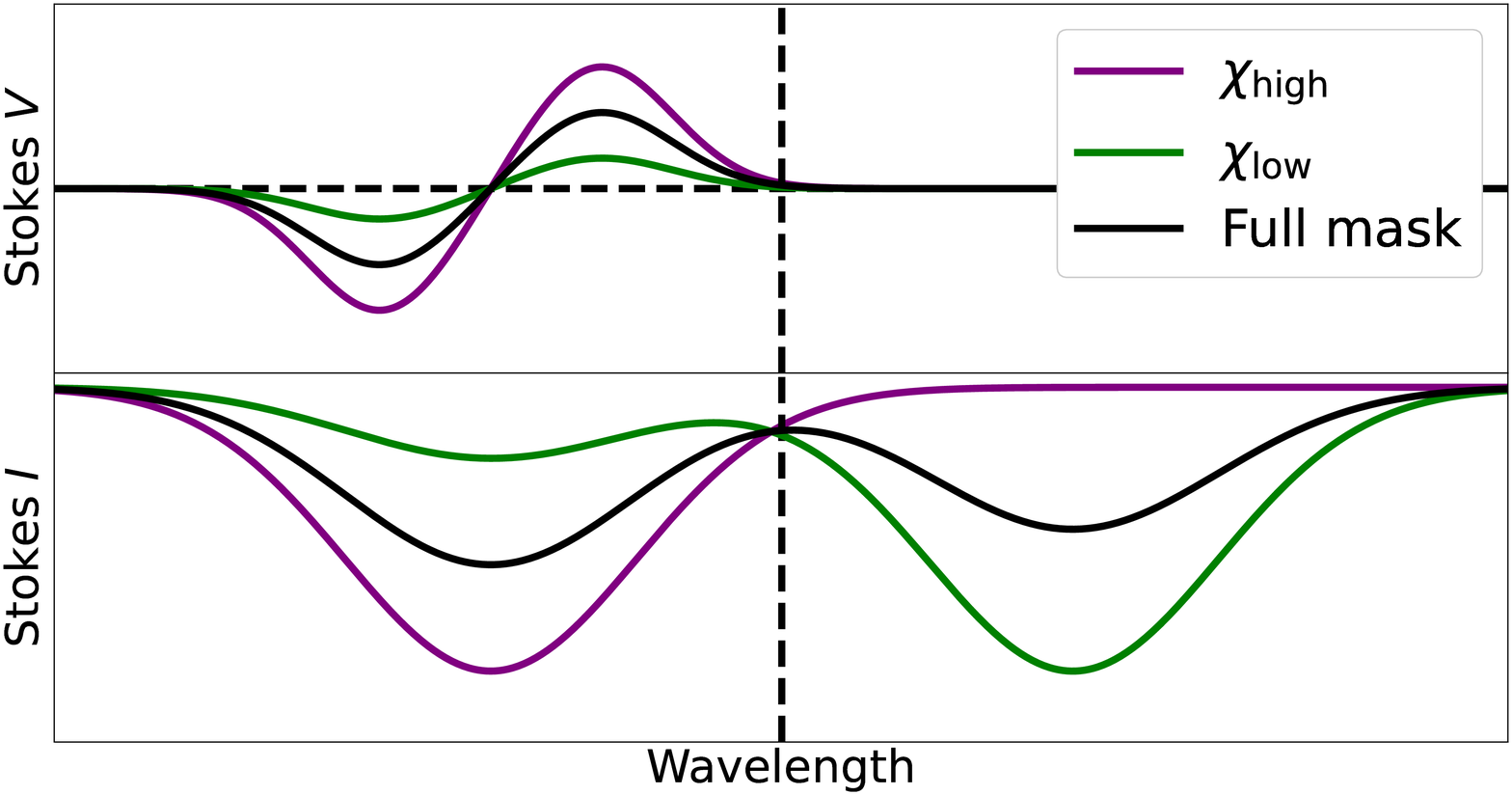}
	\caption[]{Upper panel: a schematic representation (not up to scale) of two global motions in the atmosphere of R~Sct -- an upward one in the near-photospheric layers, caused by the passing of a radiative shockwave, and a downward one in the higher atmosphere, caused by the gravitational infall occuring after the propagation of a  previous shockwave. Bottom panel: a sketch of the LSD profiles that are expected to be obtained using a mask that is restricted to lines with high $\chi$ which probe only the near-photospheric layers of the atmosphere (purple), low $\chi$ which probe the same layers plus the higher atmosphere (green), and a mask which contains a mixture of both (black).}
	\label{fig:rsct_chi_layers}
\end{figure}

\noindent This physical scenario is presented schematically in Figure~\ref{fig:rsct_chi_layers}. In the upper panel of this figure, a schematic representation of the atmosphere of R~Sct is shown (not up to scale). Periodically, a shockwave emerges from the photosphere. The narrow region just above the photosphere is affected by the shock and moves in a global upward motion. The higher atmosphere however is not yet affected by this shock; instead, it is moving down, falling ballistically after having been lifted in the past by a previous shockwave. An LSD mask $\chi_{\rm high}$ which contains only lines with excitation potential high enough to probe only the atmospheric layers lying close to the photosphere would hence yield an LSD profile showing a fully blueshifted intensity profile, since all these layers are globally moving away from the star. If there is a surface magnetic field above the detection threshold, the LSD profile would also show a Stokes~$V$ signature associated to the blueshifted intensity profile (see bottom panel in Fig.~\ref{fig:rsct_chi_layers}).\\

\noindent On the other hand, an LSD mask $\chi_{\rm low}$ containing lines with lower excitation potential which may be formed at any altitude in the atmosphere of the star would probe both the photosphere and the higher atmosphere: it would yield an LSD profile containing a mixture of information from both the layers adjacent to the photosphere (which are moving globally upwards) and the higher layers of the atmosphere (which are moving globally downwards as they are relaxing after the passage of a previous shock). This means that the intensity profile will present a double peak according to the Schwarzschild mechanism. On the contrary, as the intensity of the magnetic field decreases with altitude, a magnetic field above the detection threshold will be only observed in deep/photospheric layers, and thus associated with the blueshifted component of the Stokes~$I$ profile. Using a mask containing lines predominantly formed in the higher layers will dilute the signal and thus lead to Stokes~$V$ signature with a lower amplitude than the one obtained using the $\chi_{\rm{high}}$ mask. Finally, an LSD line mask which contains all atomic lines regardless of their excitation potential (i.e. containing all the lines in both the $\chi_{\rm high}$ and $\chi_{\rm low}$ masks) would produce an LSD profile similar to the one shown in black in the bottom panel of Fig.~\ref{fig:rsct_chi_layers}: containing information predominantly from the near-photospheric layer, since both the high- and low excitation potential lines trace the conditions in it, but also information from the higher atmosphere, which is not relevant when studying magnetism and dynamics at the photospheric level. A detailed study on the dependency of the depth of line formation on excitation potential is presented in \cite{josselinplez2007} (more specifically their Section~3.1).\\

\begin{figure} \centering
	\includegraphics[width = \linewidth]{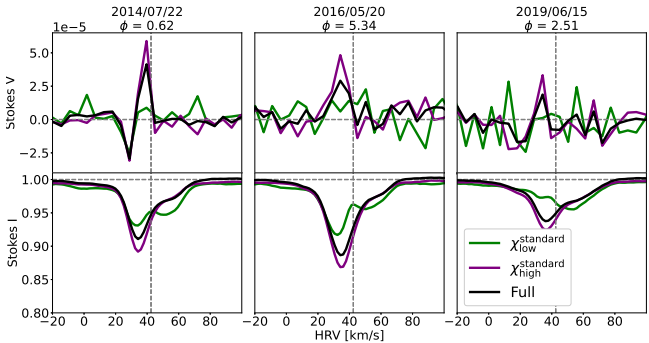}
	\caption[]{Comparison between the LSD profiles obtained with the $\chi_{\rm low}^{\rm standard}$ (containing 7903 lines, in green), $\chi_{\rm high}^{\rm standard}$ (containing 9031 lines, in purple) and full \textit{standard} (containing 16934 lines, in black) line masks for 2014/07/22 (left column), 2016/05/20 (middle column) and 2019/06/15 (right column). The Stokes~$V$ and intensity profiles are shown in the top and bottom rows respectively. The zero-level of the Stokes~$V$ and the continuum level of the intensity are indicated by the dashed horizontal lines in their corresponding panels, and the radial velocity of R~Sct is indicated by the dashed vertical line.}
	\label{fig:rsct_lsd_chi_comparison}
\end{figure}

\noindent To examine the impact lines with different excitation potential have on the LSD profiles, we performed a test by splitting the \textit{standard} LSD mask into two sub-masks: one with $\chi < 2$~eV (7903 lines) and one with $\chi \geq 2$~eV (9031 lines). These masks will hereafter be referred to respectively as $\chi_{\rm low}^{\rm standard}$ and $\chi_{\rm high}^{\rm standard}$. We picked 2~eV to be the separation limit because this value had to be high enough to successfully restrict the LSD process to the near-photospheric layers of the atmosphere and at the same time low enough to allow for a sufficient number of lines in the mask, in order to get Stokes~$V$ signatures with high SNR. We then applied the LSD method using both sub-masks to the observations obtained outside a deep minimum light and compared the results. We did the same for observations collected around deep minimum light but treated with the \textit{cool} mask instead. An example of this comparison is shown in Figure~\ref{fig:rsct_lsd_chi_comparison}. In this figure, it can be seen that the Stokes~$I$ profiles obtained with the $\chi_{\rm low}^{\rm standard}$ and $\chi_{\rm high}^{\rm standard}$ sub-masks have completely different shapes, showing that indeed the lines used for their construction probe layers of the atmosphere with different extent, and which are also affected by different dynamics, and thus should be mixed together only with great caution.
It can also be seen in Figure~\ref{fig:rsct_lsd_chi_comparison} that the lines with higher excitation potential, which are expected to trace -- at a given moment in time-- only one atmospheric dynamics at the photospheric level, show a stronger degree of circular polarization. Their associated V-signal is also blueshifted, and it appears to be linked to the blue-component of the I-profile. 
While for some observations the $\chi_{\rm high}^{\rm standard}$ sub-mask yields LSD profiles with clear Stokes~$V$ signatures, the $\chi_{\rm low}^{\rm standard}$ sub-mask never produces any LSD profiles with a signature in Stokes~$V$. From this figure, one can appreciate the level of contamination low excitation potential lines introduce to the signal originating from the photosphere.\\

\noindent The Stokes~$I$ and Stokes~$V$ profiles obtained using the $\chi_{\rm high}^{\rm standard}$ sub-mask are similar in shape to the ones obtained using the full \textit{standard} mask. This is in agreement with the scenario presented in Figure~\ref{fig:rsct_chi_layers}, since it shows that while both lines with low and high excitation potentials (which together constitute the full \textit{standard} LSD mask) trace the photospheric layer, only the higher excitation potential ones are limited strictly to it, while lines with lower $\chi$ also contain information from higher atmospheric layers (and hence unrelated to the surface magnetic field). It must be noted that the LSD profiles we obtain using the $\chi_{\rm low}$ submasks (both \textit{standard} and \textit{cool}) show no trace of Stokes~$V$ signatures in any observation. This means that the extended atmosphere of R~Sct (probed by low excitation potential lines) as a whole does not show traces of a magnetic field, but only its bottom layers (probed by high excitation potential lines) do, which is a further justification of the restriction $\chi \geq 2$~eV we use in our LSD analysis.\\

\noindent From what is seen in Figure~\ref{fig:rsct_lsd_chi_comparison}, it can also be concluded that while the photospheric layers of R~Sct show traces of a variable magnetic field, the atmosphere of the star as a whole (represented by the $\chi_{\rm low}^{\rm standard}$ sub-mask) does not. Furthermore, for the observations where the $\chi_{\rm high}^{\rm standard}$ one shows clear Stokes~$V$ signatures, the LSD profiles obtained with the $\chi_{\rm low}^{\rm standard}$ sub-mask remain featureless. One possible reason for this may be a dilution effect of the magnetic field in the upper atmosphere. 
Whatever the reason for it may be, the lack of Stokes~$V$ signatures in the LSD profiles obtained from lines with low excitation potential is a proof that mixing together atomic lines formed in different atmospheric layers can have a serious impact when investigating surface magnetism in pulsating stars hosting extended atmospheres wherein shockwaves propagate. The valuable result from this new approach for the analysis of the surface magnetic field is then a better definition of the Stokes~$V$ profiles which originate at the level of the photosphere.\\

\noindent Considering what has been said so far, the $\chi_{\rm high}^{\rm standard}$ sub-mask appears to be a better choice when analysing the magnetic field at the surface level. The exact same logic applies to the observations obtained near a deep minimum light, for which we constructed a $\chi_{\rm high}^{\rm cool}$ mask by extracting from the \textit{cool} mask the lines with $\chi \geq 2$~eV.

\subsection{The origin of circular polarization}
\label{sec:rsct_surface_field}

\begin{figure*}
    \includegraphics[width = \linewidth]{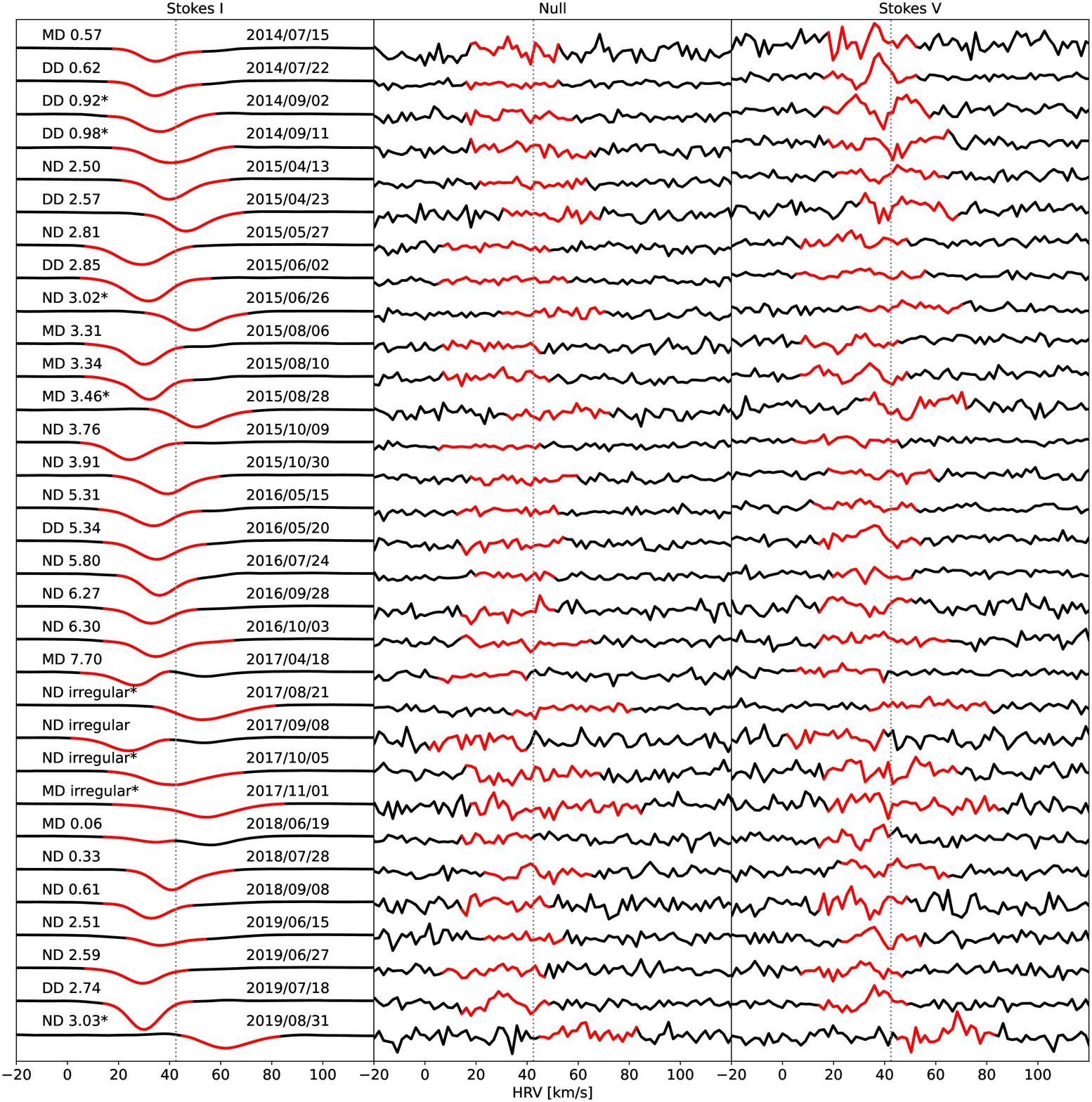}
    \caption[]{LSD profiles of R~Sct in circular polarization using only atomic lines with $\chi \geq 2$~eV. The Stokes~$I$, null and Stokes~$V$ profiles of the observations are stacked vertically in the left, middle and right column, respectively, with the date phase and detection type indicated just above their corresponding Stokes~$I$ profile. Observations treated with the \textit{cool} mask are noted with a symbol  (*). The signal (in Stokes~$I$, null and Stokes~$V$) within the interval of computation of the longitudinal magnetic field is colored in red. In each panel, the radial velocity of the star $v_{\rm rad}~=~42.47$~km~s$^{-1}$ \citep{kunder2017} is indicated with a dashed vertical gray line.}
    \label{fig:rsct_lsd_high_chi}
\end{figure*}

\noindent Using respectively the $\chi_{\rm high}^{\rm standard}$ or $\chi_{\rm high}^{\rm cool}$ mask, we applied this refined LSD method to the Narval observations of R~Sct in circular polarization in order to extract the mean polarized signatures (Stokes~\textit{V} profiles). The resulting LSD profiles are shown in Figure~\ref{fig:rsct_lsd_high_chi} and discussed in Section~\ref{sec:rsct_chi_results}.
When present, the mean Stokes~$V$ signatures are typically weak, of the order of $5 \times 10^{-5}$ the intensity of the unpolarized continuum. Clear Stokes~$V$ signatures are found for example in July and September 2014, in April and August 2015, in May 2016 and in June 2018. In some observations however, there is no trace of a Stokes~$V$ signature, for example in September 2016 and in April and August 2017.\\

\noindent Since we are dealing with faint polarized signatures obtained with Narval, an effect that must be taken into consideration is \textit{cross-talk}, or the contamination of circular polarization by linear polarization and vice versa. In practice, this effect is most notable in stars that have strong net linear polarization and weak circular polarization in their spectrum (as it is the case for R~Sct, see Section~\ref{sec:rsct_linear}). The cross-talk on Narval has been measured from regular observations of the Ap star $\gamma$~Equ. It has been estimated to be of the order of 3\%  from 2009 observations \citep{silvester2012}, and stabilized below 1.5\% from 2016 observations \citep{mathias2018}. Furthermore, \cite{tessore2017} performed a dedicated observational procedure on the red supergiant $\mu$~Cep (exhibiting Stokes U\&Q signatures stronger than Stokes V signals).  From 2015 to 2017 observations, these authors could model the cross-talk, and they obtained respectively 3.6\% and 1.4\%, for $Q$ to $V$ and $U$ to $V$,  i.e. a cross-talk level consistent with the one obtained from the Ap star $\gamma$~Equ (those results also confirmed the cross talk reciprocity between linear and circular polarization signals). So in the epoch of our R~Sct observations (2014-2019), we can safely consider the cross-talk level of Narval (from linear to circular, and reciprocally) was at most of the order of 3\%. In the following, we have considered a cross-talk level of 3\%.\\

\begin{figure*} \centering
    \includegraphics[width = \linewidth]{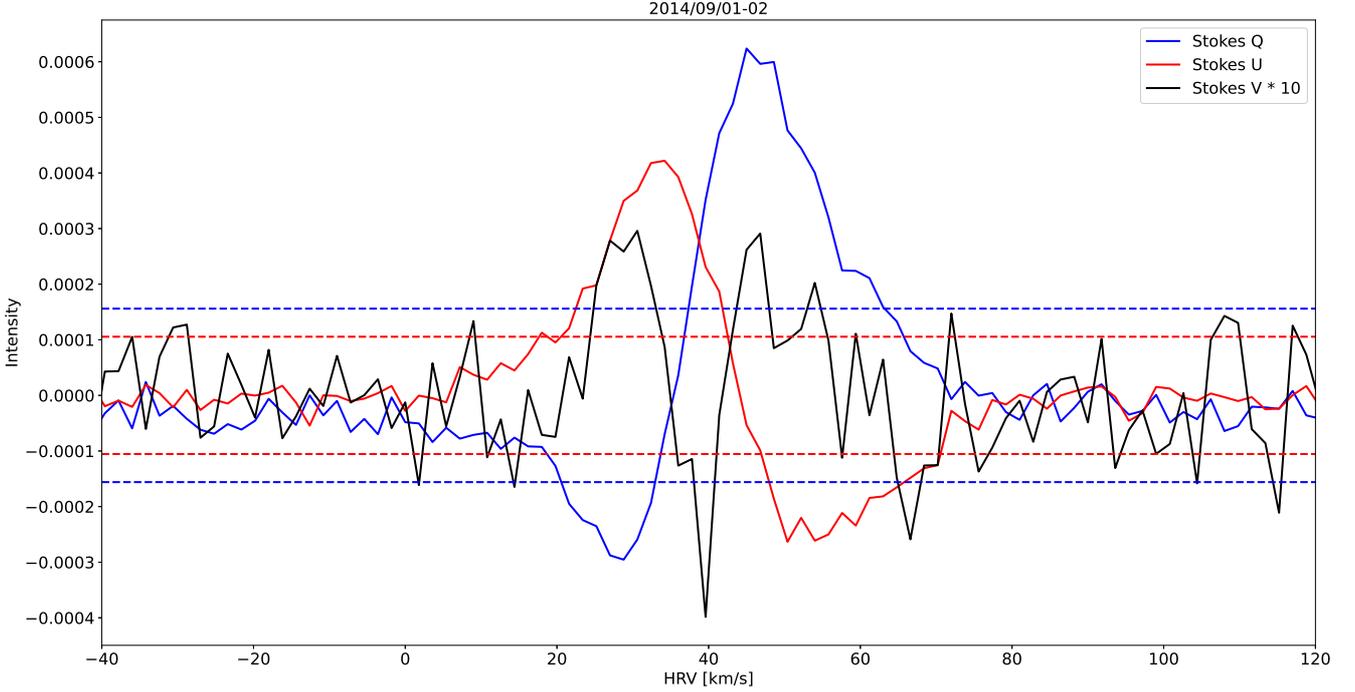}
	\caption[]{LSD profiles of R~Sct: Stokes~$Q$ (solid blue) and $U$ (solid red), obtained on 2014/09/01, and Stokes~$V$ (solid black, magnified 10 times) obtained on 2014/09/02. The horizontal dashed blue and red lines indicate the maximum expected level of pollution due to crosstalk by the Stokes~$Q$ and $U$ signal, respectively, which corresponds to 3\% of the maximum signal. This 3\% level is also magnified 10 times in the figure in order to compensate the magnification of the Stokes~$V$ signal. It can be seen that the circularly polarized profile is above the maximum expected crosstalk level, with the three peaks at 28, 40 and 45 km/s being respectively 1.6, 2.1 and 1.5 times above it, ruling out the possibility that the Stokes~$V$ signal is measured due to crosstalk from linear polarization.}
	\label{fig:rsct_crosstalk}
\end{figure*}

\noindent In order to check if the Stokes~$V$ signatures observed in R~Sct may be due to cross-talk from linear polarization, we compared the Stokes~$U$, $Q$ and $V$ LSD profiles obtained on 2014/09/01-02. These exact observations were used because 1) they all show strong and clear signatures in their respective Stokes parameter, and 2) they were made very close in time to each other. This comparison is shown in Figure~\ref{fig:rsct_crosstalk}, where the Stokes~$V$ signal is amplified 10 times in order to be more easily comparable to the much stronger $U$\&$Q$ signals. In this figure, we also show the maximum expected level of cross-talk contribution from both the Stokes~$Q$ (in dashed blue) and $U$ (in dashed red), which is below the level of the Stokes~$V$ signature observed. This shows that the observed circular polarization signal cannot be due to cross-talk, and must instead be of stellar origin.\\

\noindent To test if the Stokes~$V$ signal found in the LSD profiles of R~Sct is due to the Zeeman effect, we performed three tests using LSD with different line masks, as it has been proposed by \cite{mathias2018}. The tests aimed to check if:
\begin{enumerate}
\item{Spectral lines with high effective Land{\'e} factors ($g_{\rm eff}$) show stronger Stokes~$V$ signatures than those with low $g_{\rm eff}$;}
\item{Strong lines (ones that have a central depth larger than the average) show stronger Stokes~$V$ signatures than weak ones do;}
\item{Spectral lines formed at longer wavelengths show stronger Stokes~$V$ signatures than those formed at shorter wavelengths.}
\end{enumerate}

\begin{figure*} \centering
    \includegraphics[width = \linewidth]{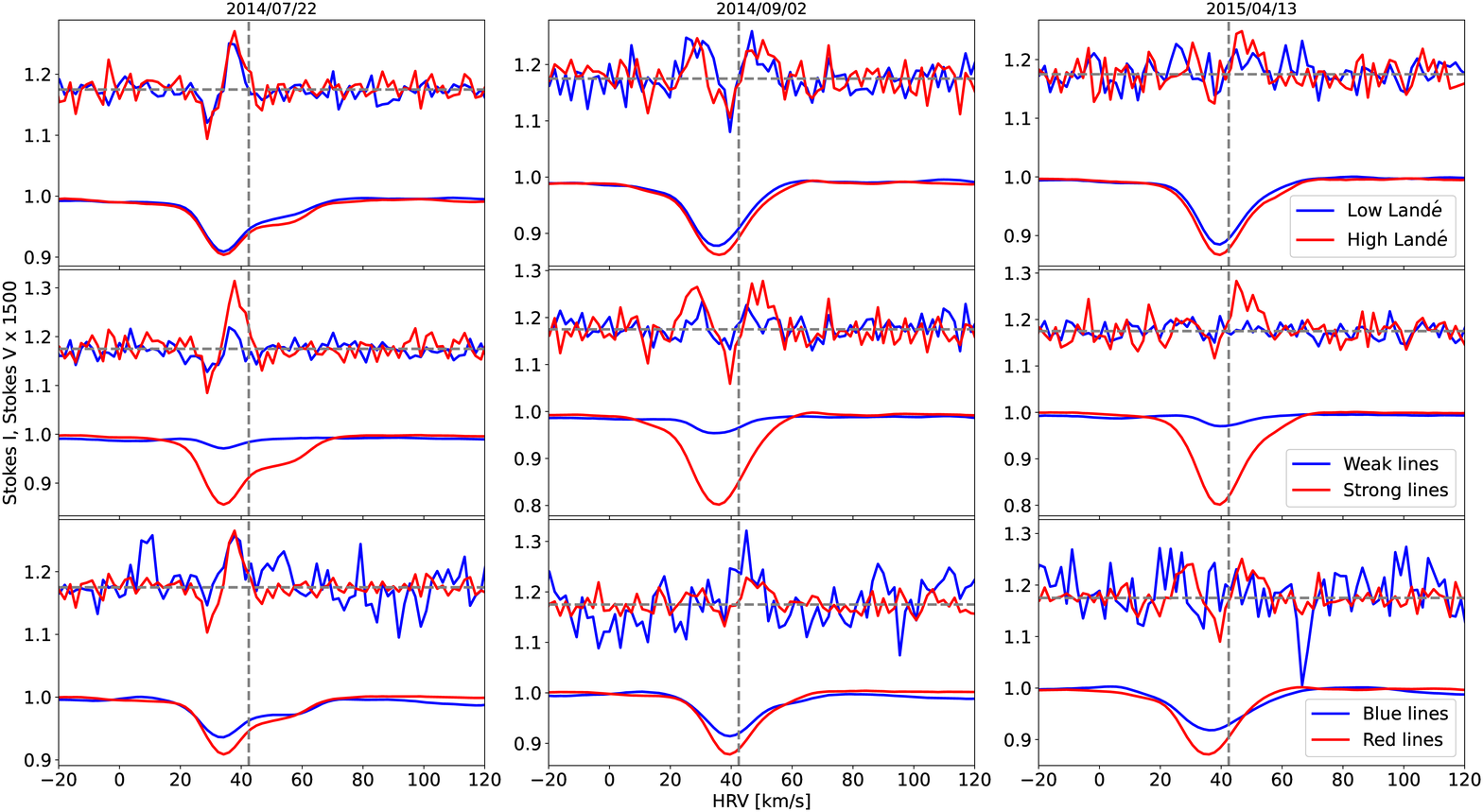}
	\caption[]{Comparison of a sample of LSD profiles of R~Sct computed with different sub-masks to test (through the Land{\'e} factor $g_{\rm eff}$, line depth and wavelength in the left, middle and right column, respectively) the Zeeman origin of the Stokes~$V$ signal (see Section~\ref{sec:rsct_surface_field}). Each column represents Stokes~$V$ profiles computed with three different pairs of submasks for a single date of observation: 2014/07/22 is obtained just after a maximum visual brightness, 2014/09/02 corresponds to a deep minimum, and 2015/04/13 is obtained between a maximum and a shallow minimum. For each observation, the radial velocity of the star $v_{\rm rad}~=~42.47$~km~s$^{-1}$ \citep{kunder2017} is indicated with a dashed vertical gray line. The Stokes~$V$ profiles are shifted vertically and amplified 1500 times for display purposes, and their zero-level is indicated with a horizontal dashed gray line.}
	\label{fig:rsct_zeeman_tests}
\end{figure*}

\noindent To perform the first test, each of the two line masks ($\chi_{\rm high}^{\rm standard}$ and $\chi_{\rm high}^{\rm cool}$) was split into two sub-masks of high and low $g_{\rm eff}$ using the median value $g_{\rm eff} = 1.2$ as the limit. For the observations where a Stokes~$V$ signature was already found using respectively the $\chi_{\rm high}^{\rm standard}$ or $\chi_{\rm high}^{\rm cool}$ mask, we applied the LSD process using the respective sub-masks for high and low $g_{\rm eff}$ and then compared the results. We found that the degree of circular polarization is indeed stronger for lines with high $g_{\rm eff}$. We performed the other two tests in a similar manner, using the median values of line depth (0.66) and wavelength (446 nm) as limits. The line depth test also showed that strong lines are more strongly polarized in Stokes~$V$ than weak ones, which according to \cite{mathias2018} favours the Zeeman origin of the circularly polarized signal. However, the wavelength test was inconclusive as it did not yield any well-defined circularly polarized signatures. Examples of the results of the tests are shown in Figure~\ref{fig:rsct_zeeman_tests}.\\

\noindent Based on these results and considering that the circularly polarized signals are of stellar origin, we conclude that the observed Stokes~$V$ profiles are of Zeeman origin and that they indicate the presence of a surface magnetic field in R~Sct.

\subsection{Magnetism at the photospheric level}
\label{sec:rsct_chi_results}

\noindent To constrain the analysis of the LSD profiles, and more specifically, the Stokes~$V$ signatures, to the level of the photosphere, we computed the LSD profiles for the Narval observations in circular polarization using the $\chi_{\rm high}^{\rm standard}$ and $\chi_{\rm high}^{\rm cool}$ masks, the latter being used for the observations obtained close to a deep minimum where TiO bands are present in the spectra. These masks were built as described in Section~\ref{sec:rsct_line_selection_chi} and the resulting LSD profiles are shown in Figure~\ref{fig:rsct_lsd_high_chi}. 
Generally, the LSD profiles obtained for $\chi \geq 2$~eV resemble those obtained using the full \textit{standard} and \textit{cool} masks. The Stokes~$V$ profiles obtained by using only high excitation potential lines however are more well-defined. This result is in agreement with the assumption of the zones of formation described in Section~\ref{sec:rsct_line_selection_chi}. It also shows that the magnetic field at the surface level differs from that in the higher atmosphere, otherwise the Stokes~$V$ profiles presented in Figure~\ref{fig:rsct_lsd_high_chi} and the ones we obtain with the full line masks would be identical (see Fig.~\ref{fig:rsct_lsd_chi_comparison}). For some particular dates (2015/08/28, 2016/05/20) it is also possible to notice a signature when using the $\chi_{\rm high}$ masks, while it is not apparent with the full ones. 
In Figure~\ref{fig:rsct_lsd_high_chi} it can be seen that for some observations (e.g. 2014/07/22, 2016/05/20) where the LSD intensity profile shows a double-peak structure, the Stokes~$V$ signature is clearly only associated to the blueshifted lobe. We propose that this magnetic field detection may be indeed associated to an ascending global motion of the photosphere caused by a radiative shock wave propagating outwards. As previous studies of cool pulsating stars by \cite{lebre2015} and \cite{sabin2015} have suggested, the propagation of a shock wave may locally amplify a weak surface magnetic field due to a compression of the magnetic field lines.\\

\noindent The fact that in Figure~\ref{fig:rsct_lsd_high_chi} a few observations still show a double-peak Stokes~$I$ profile (e.g., 2017/04/18 and 2017/09/08) indicates that the $\chi_{\rm high}$ mask does not fully constrain the LSD process to the near-photospheric layers (otherwise all LSD intensity profiles would have the shape of a single gaussian). This most likely means that the 2~eV limit may be too low, and a higher threshold is necessary in order to fully isolate the contribution of the near-photospheric layers in the computation of the LSD profiles. However, setting the excitation potential limit higher would further limit the number of lines in the mask, which would in turn decrease the SNR of the LSD profiles.\\

\noindent From Figure~\ref{fig:rsct_lsd_high_chi}, a rough estimation can be made of the typical time-scale of Stokes~$V$ variability. Let us consider the data from 2015 (set 2). There, a signal appears to be present in the beginning of April, which is not visible in May and June. It then reappears in August and vanishes again in October. From this variation, a rough time-scale of 2-3 months can be deduced, which is similar to the shock apparition time-scale (whatever main or secondary shock). This 2-3 month time-scale does not contradict the rest of the whole dataset: for example the two observations of September 2014 show that the Stokes~$V$ signal may persist for at least one week. On the other hand, the signal obviously changes quicker than on time-scales $\geq 1$~yr.\\

\begin{table}
    \centering
    \begin{tabular}{llrrrr}
Date                    & HJD    & Phase & Detection          & $B_l$ {[}G{]}      & ${\sigma}$ {[}G{]} \\
                        &        &       & ($\chi \geq 2$ eV) & ($\chi \geq 2$ eV) & ($\chi \geq 2$ eV) \\ \hline
2014/07/15              & 6854.5 & 0.57  & MD                 & -1.41              & 1.28               \\
2014/07/22              & 6861.4 & 0.62  & DD                 & -0.56              & 0.36               \\
2014/09/02$^{\ddagger}$ & 6903.3 & 0.92  & DD                 & -0.48              & 0.56               \\
2014/09/11$^{\ddagger}$ & 6912.3 & 0.98  & DD                 & -1.84              & 0.71               \\
2015/04/13              & 7126.6 & 2.50  & ND                 & -0.18              & 0.52               \\
2015/04/23              & 7136.6 & 2.57  & DD                 & 1.18               & 0.65               \\
2015/05/27              & 7170.5 & 2.81  & ND                 & -0.20              & 0.50               \\
2015/06/02              & 7176.6 & 2.85  & DD                 & 0.20               & 0.29               \\
2015/06/26$^{\ddagger}$ & 7200.5 & 3.02  & ND                 & -0.08              & 0.37               \\
2015/08/06              & 7241.9 & 3.31  & MD                 & -0.57              & 0.46               \\
2015/08/10              & 7245.4 & 3.34  & MD                 & 1.01               & 0.60               \\
2015/08/28$^{\ddagger}$ & 7263.4 & 3.46  & MD                 & -1.55              & 0.84               \\
2015/10/09              & 7305.3 & 3.76  & ND                 & 0.11               & 0.35               \\
2015/10/30              & 7326.3 & 3.91  & ND                 & 0.61               & 0.49               \\
2016/05/15              & 7524.6 & 5.31  & ND                 & 0.39               & 0.43               \\
2016/05/20              & 7529.6 & 5.34  & DD                 & 0.55               & 0.53               \\
2016/07/24              & 7594.4 & 5.80  & ND                 & 0.45               & 0.29               \\
2016/09/28              & 7660.3 & 6.27  & ND                 & 0.32               & 0.69               \\
2016/10/03              & 7665.3 & 6.30  & ND                 & 0.90               & 0.76               \\
2017/04/18              & 7862.6 & 7.70  & MD                 & 0.45               & 0.52               \\
2017/08/21$^{\ddagger}$ & 7987.4 & --    & ND                 & -0.53              & 0.58               \\
2017/09/08              & 8005.3 & --    & ND                 & 0.70               & 1.30               \\
2017/10/05$^{\ddagger}$ & 8032.3 & --    & ND                 & 0.11               & 1.24               \\
2017/11/01$^{\ddagger}$ & 8059.8 & --    & MD                 & 3.21               & 1.31               \\
2018/06/19              & 8289.5 & 0.06  & MD                 & -2.91              & 1.46               \\
2018/07/28              & 8328.5 & 0.33  & ND                 & 0.89               & 0.59               \\
2018/09/08              & 8370.4 & 0.61  & ND                 & 0.37               & 0.91               \\
2019/06/15              & 8650.5 & 2.51  & ND                 & 0.76               & 0.78               \\
2019/06/27              & 8662.5 & 2.59  & ND                 & -0.22              & 0.76               \\
2019/07/18              & 8683.5 & 2.74  & DD                 & -1.58              & 0.35               \\
2019/08/31$^{\ddagger}$ & 8727.4 & 3.03  & ND                 & -2.83              & 1.28              
    \end{tabular}
    \caption{Log of Stokes~$V$ observations and associated $B_l$ values. Calendar and heliocentric Julian dates (HJD) are given in the first two columns. HJD starts from 2~450~000. The third column lists the phase calculated from Equations~\ref{eq:rsct_eph1} and \ref{eq:rsct_eph2}. The fourth column indicates the type of detection : definite detection (DD), marginal detection (MD) or no detection (ND) (see Section~\ref{sec:detection}). The last two columns present the $B_l$ value and its errorbar, both computed from the $\chi_{\rm high}$ LSD profiles. The dates for which the $\chi_{\rm high}^{\rm cool}$ mask is used are indicated with the symbol $\ddagger$, and for the rest, the $\chi_{\rm high}^{\rm standard}$ mask is used (see Section~\ref{sec:rsct_masks_teff}).}
    \label{tab:log_stokesv}
\end{table}

\noindent From the LSD profiles shown in Figure~\ref{fig:rsct_lsd_high_chi}, we computed the longitudinal component $B_l$ using the \textit{specpolFlow} package that uses the first order moment method \citep{donati97}. For those observations that show a double-peak Stokes~$I$ profile, the velocity range of the $B_l$ computation was limited so that only the blue peak is selected. The velocity range of $B_l$ calculation for each observation can be seen in red in Figure~\ref{fig:rsct_lsd_high_chi}. The resulting $B_l$ values, along with their error of estimation, are presented in Table~\ref{tab:log_stokesv}.\\

\begin{figure*} \centering
	\includegraphics[width = \linewidth]{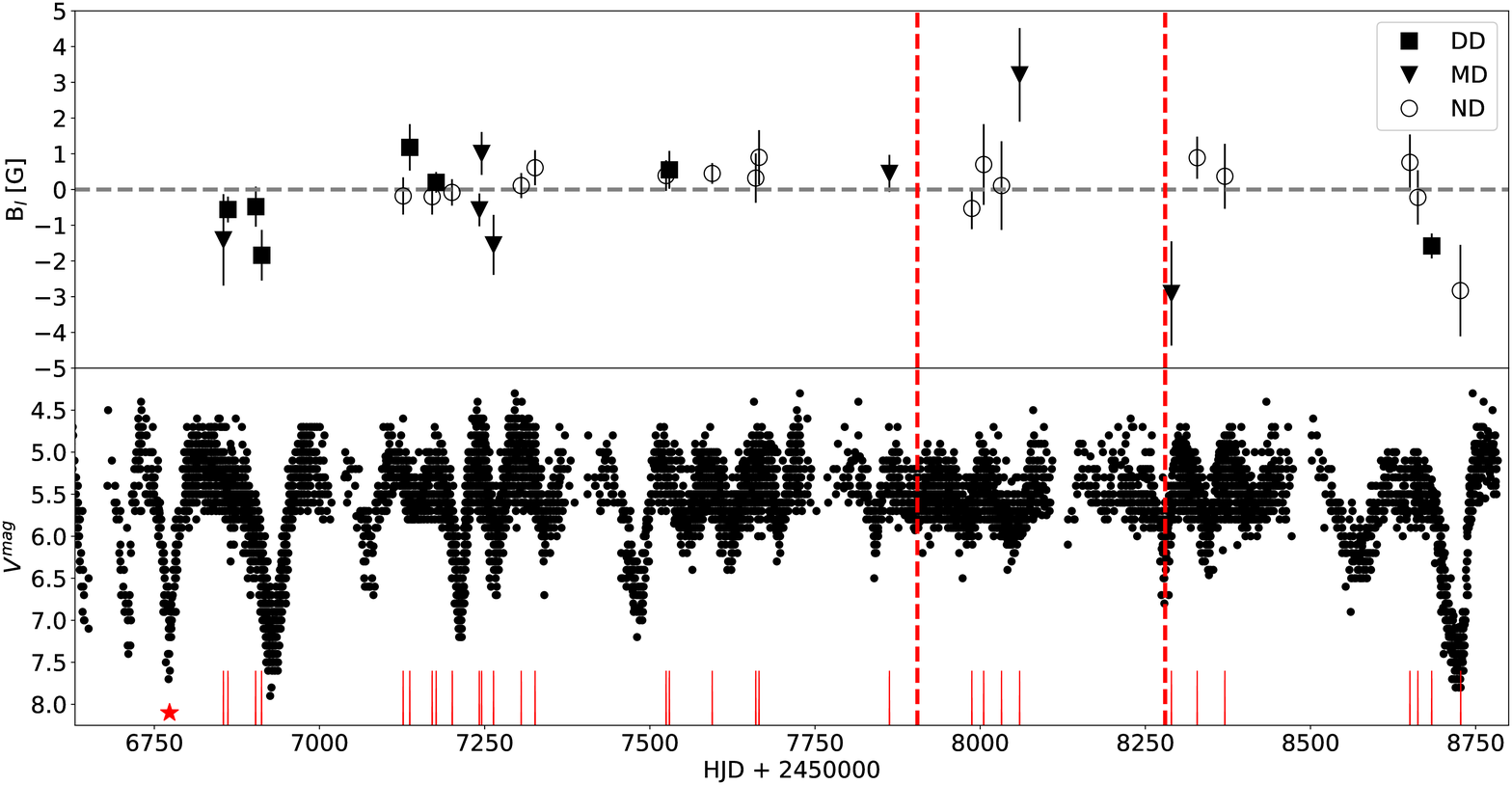}
	\caption[]{Temporal evolution of the longitudinal component of the magnetic field $B_l(t)$  (upper panel) and lightcurve (bottom panel) of R~Sct. In the upper panel, the different symbols represent the different types of detections, described in Section~\ref{sec:detection}: DD - definite detection, MD - marginal detection, ND - no detection. The red vertical lines in the bottom panel indicate the time of Narval observations that contain Stokes V measurements, and the red star indicates the date considered as the start of period~0 (see Section~\ref{sec:rsct_obs}). The dashed vertical red lines indicate the approximate beginning and end of the irregular part of the lightcurve.}
	\label{fig:rsct_bl}
\end{figure*}

\noindent The temporal evolution of the longitudinal magnetic field $B_l(t)$  is presented in Figure~\ref{fig:rsct_bl} together with the visual lightcurve. It can be seen in this figure that the strength of the longitudinal magnetic field varies in time. Looking back at Figure~\ref{fig:rsct_lsd_high_chi}, it can also be seen that the Stokes~$V$ profiles vary in position relative to the intensity profile.\\

\noindent The first half of the dataset, before the irregular portion of the lightcurve (delimited by the dashed red vertical lines in Figure~\ref{fig:rsct_lc}), contains 20 observations, of which 11 show a detection in Stokes~$V$ (6 of which are definite). The $B_l$ varies roughly between -1.8~G (2014/09/11) and 1.2~G (2015/04/23) in this first part of the observational dataset. For 3 dates -- 2014/07/22, 2015/08/06, 2016/05/20 -- the Stokes~$V$ signature appears to be associated only to the blueshifted lobe of the intensity profile. During the irregular part of the lightcurve, none of the 4 available observations show a clear polarized signature in Stokes~$V$, except for a marginal detection on 2017/11/01.
In the last part of the dataset, after the irregular portion of the lightcurve, only 2 out of 7 observations show signatures in Stokes~$V$: 2018/06/19 (MD) and 2019/07/18 (DD). Some of these results should however be considered with caution. Indeed, the standard FAP analysis may be inappropriate for weak signals such as those considered here. As recalled by \cite{2017A&A...604A.115V}, "it does not provide the probability that a given detection is spurious but rather the probability that a generic peak due to the noise in the filtered signal can exceed, by chance, a fixed detection threshold.". For example, the "non-detection" on 2019/08/31 may be regarded by naked-eye inspection as a tentative detection.\\

\noindent For those observations where the Stokes~$I$ LSD profile has a double-peak profile and a Stokes~$V$ signature is present, the latter is always associated to the blueshifted component of the Stokes~$I$ profile. This suggests that the detected magnetic field is associated to the material being lifted upwards by the passing shockwave, in support of the hypothesis that the magnetic field lines are being compressed in the shock front, which locally amplifies the surface magnetic field. However, no clear correlation exists between the longitudinal magnetic field strength and the photometric phase of the star. This suggests that the observed variability of the longitudinal magnetic field is not uniquely due to amplification caused by the propagation of shockwaves, but is also a product of a non-homogeneous and variable magnetic structure at the surface level of the star.

\section{Linear polarization}
\label{sec:rsct_linear}

\begin{figure*} \centering
	\includegraphics[width = \linewidth]{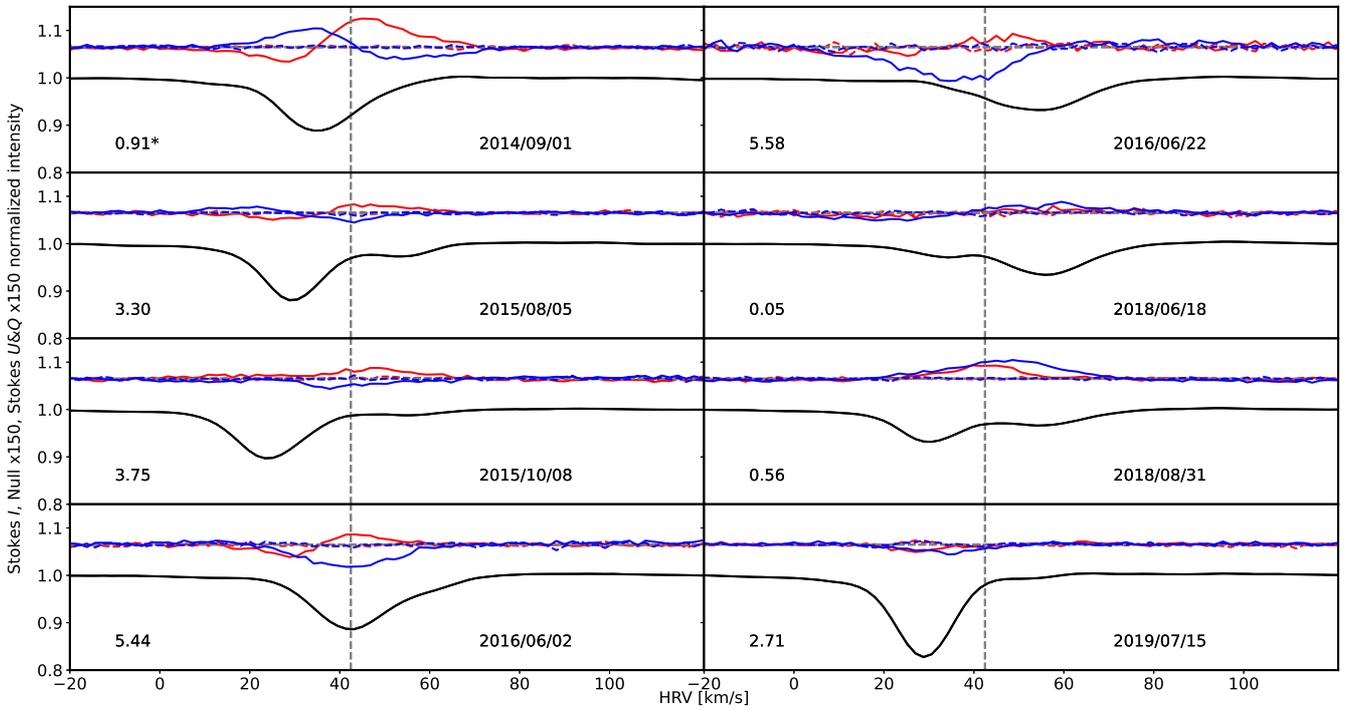}
	\caption[]{Some LSD profiles of R~Sct in linear polarization. The phase and date of observation are indicated for each panel, as in Figure~\ref{fig:rsct_lsd_high_chi}. Stokes~$Q$ profiles are shown in red and Stokes~$U$ profiles -- in blue, and their respective null profiles are shown in the same colors with dashed lines. Both the polarized $Q$\&$U$ profiles and their corresponding null profiles are shifted vertically and magnified 150 times for display purposes.}
	\label{fig:rsct_uq}
\end{figure*}

\noindent We computed the LSD profiles of the Narval observations in linear polarization listed in Table~\ref{tab:log_observations} using the full \textit{standard} and \textit{cool} masks as described in Section~\ref{sec:rsct_masks_teff} (without discriminating atomic lines by their excitation potential). We find clear linearly polarized signatures in almost all of our Narval observations, confirming the first detection of such profiles reported previously by \cite{lebre2015}. When present, these signatures usually have amplitudes of about 1~--~2~$\times 10^{-4}$ of the level of the unpolarized continuum (about 10 times stronger than the observed Stokes~$V$ signatures). A few examples of the obtained LSD profiles are presented in Figure~\ref{fig:rsct_uq}. In this figure, Stokes~$Q$ and $U$ signatures which vary in both shape and amplitude can be appreciated. It can also be seen that the signatures vary in position with respect to the Stokes~$I$ profile. For example, on 2014/09/01 the signal appears more or less centered on the Stokes~$I$ profile, while on 2015/10/08 it appears redshifted, and on 2016/06/22 it is blueshifted.\\\\

\noindent Net linear polarization within atomic spectral lines has been discovered in red supergiant (RSG) stars. For example, \cite{auriere2016} have detected the presence of Stokes~$Q$ \& $U$ signatures in the spectrum of Betelgeuse, and have shown that the origin of these signatures is depolarization of the continuum in the spectral lines. According to these authors, this is possible due to the presence of surface brightness inhomogeneities in Betelgeuse, which they suspect to be giant convective cells.

\noindent Linearly polarized signatures have also been detected by \cite{lebre2014} in another cool evolved low mass star which is known to host strong pulsations, the Mira variable $\chi$~Cyg. \cite{arturo2019} showed that the Stokes~$Q$ \& $U$ spectra of $\chi$~Cyg is dominated by intrinsic polarization, with only a negligible contribution from continuum depolarization. \cite{arturo2019} also demonstrated that the linear polarization in $\chi$~Cyg can only be explained by asymmetric velocity fields caused by the shocks, which locally enhance the Stokes~$Q$ \& $U$ signatures.\\\\

\noindent The origin of linear polarization in the spectrum of R~Sct may or may not be similar to that of $\chi$~Cyg or Betelgeuse. Further insight on the nature of linear polarization in R~Sct may be gained by detailed inspection of the polarized signal found in individual spectral lines, such as, for example, the sodium D-doublet (see e.g. \citealt{Stenflo96}). In the present work, the linearly polarized LSD profiles are only used in order to check for cross-talk between linear and circular polarization in the available Narval observations (see Section~\ref{sec:rsct_surface_field}), in order to confirm the stellar origin of the Stokes~$V$ signal. 
The analysis of the Stokes~$Q$ and $U$ profiles of R~Sct, and their origin, is to be done in a future study not in the scope of this paper. 

\section{Summary}
\label{sec:rsct_summary}

\noindent In this work, the longest monitoring of the RV~Tauri variable star R~Sct ever performed using high resolution spectropolarimetry is presented. Circular polarization signatures are detected in several observations and their stellar and Zeeman origin is confirmed. The time-scale of variation of these signatures appears to be of a few months, similar to the time-scale of the stellar pulsation. The differences between the dynamics of the lower (near-photospheric) and upper layers of the extended atmosphere of R~Sct are presented. These differences introduce the need for a refined approach for the application of the LSD method when investigating the presence of a magnetic field at the stellar surface. Such an approach is suggested and applied to the observational dataset. LSD profiles constructed from lines formed mostly in the near-photospheric layers of the atmosphere are produced, and from their Stokes~$V$ signatures, the longitudinal magnetic field is measured. When the Stokes~$I$ LSD profile shows a double-peak profile, the Stokes~$V$ signature, when present, appears to be always associated to its blueshifted lobe, suggesting a connection between the shockwave propagating outwards from the photosphere and the detection of a surface magnetic field, perhaps due to compression of magnetic field lines in the shock front.

\section*{Acknowledgements}
This work was supported by the "Programme National de Physique Stellaire" (PNPS) of CNRS/INSU co-funded by CEA and CNES. LS acknowledges support by the Marcos Moshinsky Foundation (Mexico). SG and AL thank the University of Montpellier for its support through the I-SITE MUSE project {\it MAGEVOL}. We thank F. Herpin and A. Chiavassa for their valuable comments on the text. We also thank the referee, Gregg Wade, for his valuable comments and remarks which significantly helped us improve this paper.

\section*{Data Availability}
The Narval observations used in this study are publicly available through the \textit{PolarBase} online database of spectropolarimetric observations: \url{http://polarbase.irap.omp.eu/}.\\
The AAVSO visual lightcurve can be obtained from the AAVSO website: \url{https://www.aavso.org/}.







\bsp	
\label{lastpage}
\end{document}